\documentclass{emulateapj}

\shorttitle{21-cm absorption and AGN}
\shortauthors{Curran \& Whiting}

\def\kms{km ${\rm s}^{-1}$}

\def\ch2{$\chi^2$}
\def\dg{$^{\circ}$}
\def\Mo{M$_\odot$}
\def\kms {\hbox{${\rm km\ s}^{-1}$}}

\def\scm  {$\hbox{{\rm cm}}^{-2}$}    %cm-2
\def\deg {\hbox{$^{\circ}$}} %degrees

\def\WAT {\hbox{${\rm H_{2}O}$}} %H2O

\def \AL {$\alpha $}     %  gr. alpha
\def \HI {H{\sc \,i}}

\def \WpHz {W Hz$^{-1}$}
\def\lapp{\ifmmode\stackrel{<}{_{\sim}}\else$\stackrel{<}{_{\sim}}$\fi}
\def\gapp{\ifmmode\stackrel{>}{_{\sim}}\else$\stackrel{>}{_{\sim}}$\fi}

\begin{document}


\title{\HI\ 21-cm absorption and unified schemes of active galactic nuclei}


%\author{S. J. Curran\altaffilmark{1} and M. T. Whiting\altaffilmark{2}}
\author{S. J. Curran}
\affil{School of Physics, University of New
  South Wales, Sydney NSW 2052, Australia}
\email{sjc@phys.unsw.edu.au}

\and

\author{M. T. Whiting}
\affil{CSIRO Australia Telescope National Facility, PO Box 76, Epping NSW 1710, Australia}

\begin{abstract}
In a recent study of $z\geq0.1$ active galactic nuclei (AGN), we found
that 21-cm absorption has never been detected in objects in which the
ultra-violet luminosity exceeds $L_{\rm UV}\sim 10^{23}$ \WpHz. In
this paper, we further explore the implications that this has for the
currently popular consensus that it is the orientation of the
circumnuclear obscuring torus, invoked by unified schemes of AGN,
which determines whether absorption is present along our
sight-line. The fact that at $L_{\rm UV}\lapp 10^{23}$ \WpHz, both
type-1 and type-2 objects exhibit a 50\% probability of detection,
suggests that this is not the case and that the bias against detection
of \HI\ absorption in type-1 objects is due purely to the inclusion of
the $L_{\rm UV}\gtrsim 10^{23}$ \WpHz\ sources.  Similarly, the
ultra-violet luminosities can also explain why the presence of 21-cm
absorption shows a preference for radio galaxies over quasars and the
higher detection rate in compact sources, such as CSS or GPS sources,
may also be biased by the inclusion of high-luminosity sources.  Being
comprised of all 21-cm searched sources at $z\geq0.1$, this is a
necessarily heterogeneous sample, the constituents of which have been  observed by various instruments.
By this same token, however, the dependence on the UV luminosity 
may be an all encompassing effect, superseding the unified schemes model,
although there is the possibility that the exclusive 21-cm non-detections at high UV luminosities
could be caused by a bias towards gas-poor ellipticals. Additionally, the high
UV fluxes could be sufficiently exciting/ionising the \HI\ above 21-cm
detection thresholds, although the extent to which this is related to
the neutral gas deficit in ellipticals is currently unclear.

Examining the moderate UV luminosity ($L_{\rm UV}\lapp 10^{23}$\WpHz)
sample further, from the profile widths and offsets from the systemic
velocities, we find no discernible differences between the two AGN
types. This may suggest that the bulk of the absorption generally
occurs in the galactic disk, which must therefore be randomly
orientated with respect to the circumnuclear torus. Furthermore, we
see no difference in the reddening between the two AGN types,
indicating, like the 21-cm absorption, that the orientation of the
torus has little bearing on this.  We also find a correlation between
21-cm line strength and the optical-near-infrared colour, which
suggests that the reddening is caused by dust located in the
large-scale, \HI\ absorbing disk which intervenes the sight-line to
the AGN.

\end{abstract}


\section{Introduction}

\label{intro}
Redshifted observations of the 21-cm spin-flip transition of neutral
hydrogen (\HI) trace the cool component of the gas in distant
galaxies. Since the surface brightness has a $(1+z)^4$ dependence, the
detection of the 21-cm in emission is very difficult at redshifts of
$z\geq0.1$, and so the neutral gas in distant active galactic nuclei
(AGN) is usually studied in absorption.  Furthermore, since most
published searches\footnote{Prior to the $z\sim3$ survey of
  \citet{cww+08}.}  have been at redshifts of $z\lapp1$, there are
generally no observations of the Lyman-\AL\ transition, which is
redshifted into the optical bands at $z\geq1.7$. Therefore, to date
21-cm absorption has been the most common probe of the neutral gas in
the galaxies host to AGN, being detected in approximately 40\% of
$z\gapp0.1$ cases (see \citealt{cww+08}). 

In order to explain the detection rate, many studies invoke unified
schemes of active galactic nuclei, which attempt to unify the many
classes of luminous extragalactic object, a key element of which is
the presence of a torus of highly obscuring circumnuclear material:
In these schemes, the appearance of the object
is dependent upon the orientation of this material along our
line-of-sight to the nucleus \citep{ost78,am85,mg87,ant93,up95}, with
the popular consensus being that only
type-2 objects present a dense column of intervening gas, which can
absorb in 21-cm (see \citealt{jm94,cb95}, figure 2 of
\citealt{cww08}).  Other observational examples of this include:
\begin{enumerate}
 \item From a survey for 21-cm absorption in 23 radio galaxies, \citet{mot+01}
   find that of the five detections, four occur in sources which could be
considered type-2 objects, whereas there is only one detected case for a type-1
object. 

\item From a study of 49 gigahertz peaked spectrum (GPS) and compact
  steep spectrum (CSS) sources, \citet{pcv03} find that 21-cm
  absorption is more likely to arise in objects classified as
  galaxies, rather than in quasars. Since the former are generally
  considered to be type-2 objects, while latter are
  type-1 objects, this is consistent with the orientation of the
  central obscuration playing a major r\^{o}le in producing strong
  21-cm absorption along our sight-line.

\item Also, from a study of 27 GPSs and CSSs, \citet{gss+06} find that 21-cm 
absorption is twice as likely to be detected in the galaxies 
than in the quasars of the sample, again suggesting that the absorption occurs in the
dense sub-parsec torus.

\item From a sample of 23 galaxies and 9 quasars, \citet{gs06a} find that
15 of the galaxies exhibit 21-cm absorption, compared to just a single case
for the quasars. Like \citet{pcv03}, this is consistent with unified schemes,
where galaxies are host to edge-on obscurations, whereas quasars have their
tori oriented more face-on. 
\end{enumerate}

However, the situation may be more complex than this with evidence
that 21-cm absorption may also be due to in-falling gas or outflows
(e.g. \citealt{vpt+03,mto05,mhs+07}), and
if these are directed along the radio jet axis, we would expect
outflows of neutral gas to render absorption detectable towards type-1
sources. Whether due to an outflow or the presence of an intervening
circumnuclear obscuration, these scenarios are consistent with unified
schemes of AGN, playing a major r\^{o}le in whether \HI\ 21-cm
absorption is detected.

\begin{figure*}
\centering \includegraphics[angle=270,scale=0.70]{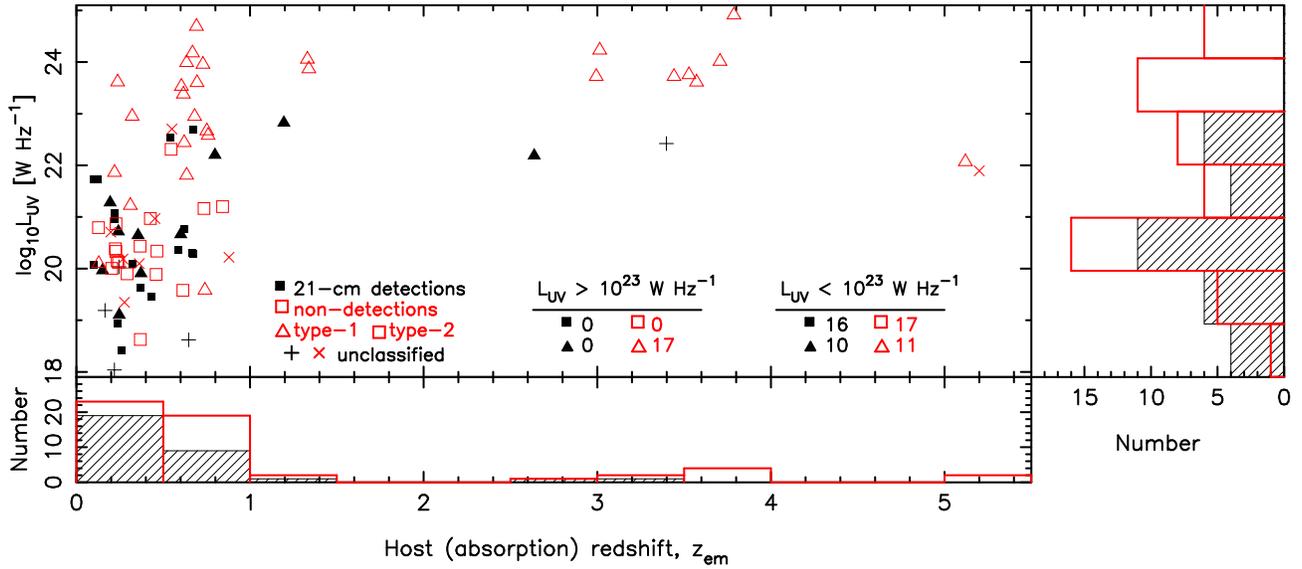}
\caption{The ultra-violet luminosity--redshift distribution for the $z\geq0.1$ radio galaxies
and quasars searched in associated 21-cm absorption. The filled symbols/hatched histogram represent the 21-cm
  detections and the unfilled symbols/unfilled histogram the
  non-detections. The shapes represent the AGN classifications, with
  triangles representing type-1 objects and squares type-2s ({\bf +}
  and {\sf x} designate an undetermined AGN type for a detection and
  non-detection, respectively). The legend shows the number of each
  AGN type according to the $L_{\rm UV}=10^{23}$
  \WpHz\ partition. Updated from \citep{cww+08}.}
\label{lum-z}
\end{figure*}

Therefore, from these possibilities in addition to absorption by
the large reservoir of neutral gas in the galactic disk, we may expect
a high 21-cm detection rate in distant radio galaxies and quasars.
However, from a recent survey of the host galaxies of $z\geq2.9$
quasars, \citet{cww+08} found no evidence of absorption in any of the
ten sources searched. Upon an analysis of the spectral types of the
targets, as well as those of all the other $z\geq0.1$ published
searches, they found the non-detections all to be type-1 objects, as
are many of the lower redshift non-detections (see
Fig.~\ref{lum-z}). 
%%%%%%%%%%%%%%% new instructions for getting numbers %%%%%%%%%%%%%
%Log luminosity cut-off for statistical tests [e.g. 21.5]? 23
% sort -n +2 -1 lum-z-class-non.dat  for L_UV column ???????
%sort -n +7 -3 low-lum.txt
%
%                           NUMBERS FOR PLOTS
%      sort -n +7 -3 low-lum-det.txt  > temp    WILL DO BY AGN TYPE
%          => 5 unknowns, 10 type-1 and 17 types-2
%      sort -n +7 -3 low-lum-non.txt > temp   => 10 unknowns, 11 type-1 and 20 types-2
%
%%%%%%%%%%%%%%%%%%%%%%%%%%%%%%%%%%%%%%%%%%%%%%%%%%%%%%%%%%%%%%%%%%%%%%%%%%%%%%%%%
Superficially, this suggests that the orientation
of the circumnuclear obscuration may be key in the detection of
21-cm absorption, although
there may also be other effects at play, the evidence for which we
discuss in this paper.

%\section{Discussion}
\section{Factors affecting the 21-cm detection rate}
%\subsection{Luminosity versus orientation effects}

\subsection{Luminosities}
\subsubsection{Ultra-violet luminosity}
\label{lum}

\begin{table*}
\begin{center}
\caption{The $z\geq0.1$ sources detected in 21-cm absorption.\label{dets}}
\small
\begin{tabular}{r c l cccc cccccc }
\tableline
Source     &Class & $z_{\rm em}$   &$B$ &$V$  &$R$ &$K$ &$\log L_{\rm UV}$ & Type & $\log_{10}\,N_{\rm HI}$ & ID & \multicolumn{2}{c}{References}\\
           &      &                &[mag]&[mag]&[mag]&[mag]&[\WpHz]& & $(f/T_{\rm s})$[\scm/K] &  & Spe. & Con.\\ 
\tableline
J0025--2602 &Gal &0.3220 &20.300  &---     &18.084  &15.674  &20.100 &2 & 18.36 & CSS & V03 & T02\\
  0108+388  &Gal &0.6685 &---     &---     &22.000  &16.690  &20.309 &2 & 19.90 & GPS & C98 & P88,B90,O98,Z02\\%J0111+3906
J0141+1353  &Gal &0.6210 &22.327  &20.920  &20.876  &16.680  &20.777 &2 & 18.04 & CSS & V03 & F89,S95\\
J0410+7656  &Gal &0.5985 &---     &---     &21.200  &---     &---    &2 & 18.40 & GPS & V03 & D95,S95a\\
J0414+0534  &Gal &2.6365 &24.100  &23.800  &21.270  &13.540  &22.188 &1 & 18.88  & --- & M99 & ---\\
J0431+2037  &Gal &0.2190 &22.174  &---     &19.085  &14.924  &18.039 &--& 18.54 & GPS & V03 & D95,S01a\\
  0500+019  &Gal &0.5846 &22.500  &21.350  &20.682  &15.430  &20.367 &2 & 18.79 & FSRS& C98 & S01b\\
   3C\,190  &QSO &1.1946 &19.976  &17.460  &18.972  &15.300  &22.825 &1 & 19.6  & CSS & I03 & ---  \\
J0834+5534  &Gal &0.2420 &18.921  &17.390  &17.180  &14.180  &20.719 &1 & 18.03 & RG  & V03 & W85\\
J0901+2901  &Gal &0.1940 &19.321  &18.078  &18.600  &15.200  &21.280 &1 & 17.04 & CSS & V03 & A95\\
  0902+343  &Gal &3.3980 &---     &23.800  &23.500  &19.900  &22.422 &--& 18.49 & --- & U91 & --- \\
J0909+4253  &QSO &0.6700 &18.960  &19.049  &18.220  &14.860  &22.699 &2 & 18.09 & CSS & V03 & V92\\
J1124+1919  &Gal &0.1650 &22.082  &21.448  &20.513  &15.930  &19.190 &--& 18.70 & CSS & G06 & S90,S95b\\
12032+1707  &Gal &0.2170 &18.758  &---     &17.327  &14.864  &20.949 &2 & 18.74 & OHM & P05 & --- \\
J1206+6413  &Gal &0.3710 &21.847  &20.790  &19.910  &---     &19.908 &1 & 18.29 & CSS & V03 & S95a,L98\\
J1326+3154  &Gal &0.3700 &21.367  &19.822  &18.882  &14.940  &19.638 &2 & 17.85 & GPS & V03 & M81,F96 \\
4C\,12.50  &QSO &0.1217 &16.615  &16.050  &15.718  &13.216  &21.736 &2 & 18.79 & GPS & M89 & L03\\
J1357+4354 &Gal &0.6460 &---     &22.708  &20.951  &---     &18.620 &--& 19.52 & GPS & V03 & T96\\
J1400+6210 &Gal &0.4310 &22.137  &20.373  &19.530  &16.130  &19.459 &2 & 18.27 & GPS & V03 & D95\\
  1413+135 &QSO &0.2467 &21.055  &20.000  &18.461  &14.928  &19.105 &1 & 19.11 & CSS & C92  & P96,P00\\
  1504+377 &Gal &0.6715 &---     &21.808  &20.800  &16.100  &20.295 &2 & 19.65 &FSRS& C98 & VLBA\\
  1549--79 &Gal &0.1501 &---     & 18.800 & ---    &12.407  &19.965 &1 & 18.56 & CFS&  M01 & same \\%\multicolumn{2}{l}{~~~~~~~~~M01} \\
J1815+6127 &QSO &0.6010 &21.272  &---     &19.122  &---     &20.665 &1 & 18.64 & GPS & V03 & T94\\
J1816+3457 &Gal &0.2448 &20.342  &---     &18.459  &15.525  &20.034 &--& 18.71 & GPS & P00 & same \\%\multicolumn{2}{l}{P00}\\
J1821+3942 &Gal &0.7980 &19.598  &---     &18.135  &15.023  &22.202 &1 & 18.22 & CSS & V03 & D95,S01a\\
J1944+5448 &Gal &0.2630 &21.732  &---     &18.591  &15.000  &18.424 &2 & 18.69 & GPS & V03 & S01a,X95\\
J1945+7055 &Gal &0.1010 &18.726  &---     &17.199  &13.369  &20.067 &2 & 18.5  & GPS & P99 & T97\\
J2052+3635 &Gal &0.3550 &22.083  &---     &21.200  &---     &20.648 &1 & 18.86 & GPS & V03 & P81 \\
3C\,433    &Gal &0.1016 &17.660  &16.350  &---     &12.891  &21.739 &2 & 18.36 & RG  & M89 & --- \\
J2255+1313 &QSO &0.5430 &19.535  &19.590  &19.190  &---     &22.530 &2 & 17.62 & CSS & V03 & A95\\
J2316+0405 &Gal &0.2199 &18.595  &17.440  &17.220  &13.991  &21.081 &2 & 17.85 & BLRG& V03 & T03\\
J2355+4950 &Gal &0.2379 &21.101  &---     &18.400  &15.112  &18.940 &2 & 18.45 & GPS & V03 & P95,T00\\
\tableline
\end{tabular}
\tablecomments{%The sources are listed by their names as given by the
  references in the penultimate column (3C\,190 = 0758+143, 4C\,12.50
  = 1345+12, 3C\,433 = 2121+24, 3C\,452 = J2245+3941).  The references
  for the magnitudes and AGN types are given in \citet{cww+08} [see
    also footnote \ref{new_sources}]. The 1216 \AA\ luminosities and
  AGN types are calculated and determined as described in
  \citet{cww+08} and the final four columns give the \HI\ column
  density, radio ID (see notes) and the 21-cm search (spe.) \& high
  resolution radio imaging (where available, con.)  references. \\BLRG
  -- broad line radio galaxy, CFS -- compact flat spectrum, CSO --
  compact symmetric object, CSS -- compact steep spectrum source, EORG
  -- end-on radio galaxy, EORQ -- end-on radio, quasar, FRI --
  Fanaroff Riley type-1/BL Lac object, FRII -- Fanaroff Riley type-2,
  FSRQ -- flat-spectrum radio quasar, FSRQ -- flat spectrum radio
  source, FSSO -- flat-spectrum symmetric object, HFP -- high
  frequency peaker galaxy, NLRG -- narrow-line radio galaxy, OHM--OH
  megamaser, RG -- radio galaxy, RQ -- radio quasar.}  \tablerefs{{\em
    Spectral}: D85 -- \citet{dsm85}, M89 -- \citet{mir89}, V89 --
  \citet{vke+89}, U91 -- \citet{ubc91}, C92 -- \citet{cps92}, C98 --
  \citet{cmr+98}, M99 -- \citet{mcm98}, P99 - \citet{ptc99}, P00 --
  \citet{ptf+00}, M01 -- \citet{mot+01}, I03 -- \citet{ida03}, P03 --
  \citet{pcv03}, V03 -- \citet{vpt+03}, P05 -- \citet{pbdk05}, C06 --
  \citet{cwm+06}, G06 -- \citet{gss+06}, GS06 -- \citet{gs06}, O06 --
  \citet{omd06}, C07 -- \citet{cwh+07}, C08 -- \citet{cww+08}.  {\em
    Continuum}: J77 -- \citet{jpr77}, M81 -- \citet{mrs81}, P81 --
  \citet{pm81}, B82 -- \citet{bod+82} U83 -- \citet{ujw83}, A85 --
  \citet{au85}, W85 -- \citet{wbw+85}, G88 -- \citet{gfgp88}, P88 --
  \citet{pr88}, F89 -- \citet{ffp+89}, S89a -- \citet{smc+89}, S89b --
  \citet{som89}, B90 -- \citet{bomd90}, S90 -- \citet{srs+90}, A91 --
  \citet{asz+91}, V92 -- \citet{vff+92}, M93 -- \citet{mbp93}, T94 --
  \citep{tvp+94}, A95 -- \citet{ag95}, D95 -- \citet{dff+95}, N95 --
  \citet{nrh95}, P95 -- \citet{pwx+95}, S95a -- \citet{sjw+95}, S95b
  -- \citet{ssl+95}, X95 -- \citet{xrp+95}, B96 -- \citet{bpf+96}, F96
  -- \citet{fcf96}, P96 -- \citet{pcsc96}, T96 -- \citet{trp96}, T97
  -- \citet{tv97}, B98 -- \citet{bgg98}, L98 -- \citet{lgs+98}, S98 --
  \citet{sk98}, O98 -- \citet{ocp98}, R99 -- \citet{rkp99}, F00 --
  \citet{ffp+00}, T00 - \citet{tmpr00}, S01a -- \citet{sjs+01}, S01b
  -- \citet{sdo+01}, B02 -- \citet{bgp+02}, T02 -- \citet{tkm+02}, Z02
  -- \citet{zrk+02}, F03 -- \citet{fpm+03}, L03 -- \citet{lkv+03}, T03
  -- \citet{tsm03}, P05 -- \citet{pkfg05}, K09 -- \citet{klm+09}, VLBA
  -- VLBA calibrator.}
\end{center}
\end{table*}

\begin{table*}
\begin{center}
\caption{As Table \ref{dets}, but for the non-detections, where the \HI\ column density
      limits are quoted at a $3\sigma$ level per channel. \label{non-dets}}
\small
\begin{tabular}{r c l cccc cccccc }
\tableline
Source     &Class & $z_{\rm em}$   &$B$ &$V$  &$R$ &$K$ &$\log L_{\rm UV}$ & Type & $\log_{10}\,N_{\rm HI}$ & ID &  \multicolumn{2}{c}{References}\\
           &      &                &[mag]&[mag]&[mag]&[mag]&[\WpHz]& & $(f/T_{\rm s})$[\scm/K] &  & Spe. & Con.\\ 
\tableline
J0003+2129  &QSO &0.4520 &21.005 & 20.580 &19.650 &---    &20.971 &--& $<19.61$  & HFP &  O06 & VLBA\\
  0035--024 &Gal &0.2197 &19.110 & 17.920 & ---   &14.494 &21.862 &1 & $<17.67$  & FRII&  M01 & VLBA\\
  0131--001 &QSO &0.8790 &23.340 & 22.500 &20.780 &16.780 &20.221 &--& $<18.28$ & --  & C06 & VLBA\\
J0157--1043 &QSO &0.6160 &17.504 & ---    &17.039 &---    &23.380 &1 & $<17.98$ & EORQ& V03 & R99\\
J0201--1132 &QSO &0.6690 &16.232 & ---    &16.073 &13.860 &24.176 &1 & $<17.80$ & EORQ& V03 & R99\\
J0224+2750  &Gal &0.3102 &19.502 & ---    &18.263 &15.250 &21.225 &1 & $<18.14$ & CSS & V03 & S95a,b\\
  0335--122 &QSO &3.4420 &21.018 & 20.110 &20.199 &17.510 &23.722 &1 & $<18.32$ & --- & C08 & VLBA\\
  0347--211 &QSO &2.9940 &20.476 & ---    &20.297 &17.900 &23.722 &1 & $<18.38$ & --- & C08 & VLBA\\
J0348+3353  &Gal &0.2430 &20.723 & ---    &19.110 &14.390 &20.121 &2 & $<18.12$ & CSS & V03 & D95\\
J0401+0036  &Gal &0.4260 &20.200 & 19.010 &18.532 &---    &20.969 &2 & $<18.03$ & EORG& V03 & N95\\
J0521+1638  &QSO &0.7590 &19.370 & 18.840 &18.480 &15.380 &22.580 &1 & $<17.65$ & CSS & V03 & F89,A91,S95a\\
  0537--286 &QSO &3.0140 &19.290 & ---    &18.789 &16.770 &24.231 &1 & $<18.45$ & FSRQ& C08 & K09\\
J0542+4951  &QSO &0.5450 &18.450 & 17.800 &17.210 &---    &22.311 &2 & $<17.45$ & CSS & V03 & L98\\
J0556--0241 &Gal &0.2350 &20.968 & ---    &19.533 &---    &20.150 &2 & $<18.77$ & GPS & V03 & P05 \\
J0609+4804  &Gal &0.2769 &21.198 & ---    &18.767 &---    &19.349 &--& $<17.79$ & EORG& V03 & N95\\
J0709+7449  &Gal &0.2921 &19.982 & ---    &17.540 &13.790 &19.898 &2 & $<18.30$ & FRII& V03 & --- \\
0723--008   &QSO &0.1273 & 17.39 & 16.57  & 15.82 &13.166 &20.793 & 2  & $<17.76$ & CFS & V89 & B96\\
J0741+3112  &QSO &0.6350 &16.517 & 16.100 &16.322 &16.100 &23.990 &1 & $<17.97$ &  GPS& V03 & S98,S01\\
J0815--0308 &Gal &0.1980 &18.490 & 16.940 &16.797 &13.858 &20.707 &--& $<18.16$ & EORG& V03 & N95\\
J0840+1312  &QSO &0.6808 &18.370 & 17.940 &17.622 &15.280 &22.947 &1 & $<17.69$ & RQ  & V03 & VLBA\\
J0913+5919  &QSO &5.1200 &---    & 23.281 &24.948 &---    &22.071 &1 & $<19.34$ & CSO & C07 & ---\\
J0924--2201 &Gal &5.2000 &---    & ---    &25.850 &---    &21.893 &--& $<18.34$ & CSO & C07 & ---\\
J0927+3902  &QSO &0.6948 &17.064 & ---    &16.486 &---    &23.603 &1 & $<17.97$ & EORG& V03 & B82\\
J0939+8315  &Gal &0.6850 &---    & ---    &20.140 &---    &---    &2 & $<17.67$ & EORG& V03 & J77\\
J0943--0819 &Gal &0.2280 &19.401 & ---    &18.100 &14.750 &20.868 &2 & $<18.08$ & GPS & V03 & B02\\
J0954+7435  &Gal &0.6950 &---    & ---    &21.700 &---    &---    &--& $<18.37$ & RG  & V03 & F00\\
J1035+5628  &Gal &0.4590 &---    & 21.244 &20.200 &---    &19.889 &2 & $<18.12$ & GPS & V03 & T94,P00\\
J1120+1420  &Gal &0.3620 &---    & 20.935 &20.100 &17.100 &20.098 &--& $<17.76$ & GPS & V03 & B98\\
J1159+2914  &QSO &0.7290 &17.489 & 18.113 &17.652 &---    &23.955 &1 & $<18.26$ & EORG& V03 & A85\\
  1228--113 &QSO &3.5280 &22.010 & ---    &19.115 &16.370 &23.754 &1 & $<18.63$ &  ---& C08 & VLBA \\
J1252+5634  &QSO &0.3210 &17.760 & 17.930 &17.660 &---    &22.949 &1 & $<17.83$ & CSS & V03 & S95a,L98\\
J1308--0950 &Gal &0.4640 &20.767 & 20.500 &18.439 &---    &20.340 &2 & $<18.11$ & CSS & V03 & T02\\
J1313+5458  &QSO &0.6130 &---    & 21.735 &20.374 &---    &19.581 &2 & $<18.23$ & RQ  & V03 & T94\\
  1351--018 &QSO &3.7070 &21.030 & 19.696 &19.277 &17.070 &24.014 &1 & $<18.41$ & --- & C08 & ---\\
  1356+022  &QDO & 1.330 &---    & 17.436 & ---   &14.537 &24.055 &1 & $<17.88$ & FSSO& D85 & VLBA\\%agpw06
J1421+4144  &Gal &0.3670 &20.496 & 19.330 &18.560 &15.910 &20.435 &2 & $<17.82$ & CSS & V03 & A95\\
J1443+7707  &Gal &0.2670 &---    & ---    &18.730 &---    &---    &2 & $<18.15$ & CSS & V03 & L98\\
  1450--338 &Gal &0.3680 &22.520 & 20.400 &19.390 &15.230 &18.629 &2 & $<17.83$ & --- & C06 & VLBA\\
  1535+004  &QSO &3.4970 &---    & ---    &---    &19.540 &---    &--& $<18.22$ & FSSO& C06 & VLBA\\%agpw06
J1540+1447  &QSO &0.6050 &17.480 & 17.000 &17.240 &13.640 &23.529 &1 & $<17.77$ & EORG& V03 & U83\\
J1546+0026  &Gal &0.5500 &19.730 & 18.900 &---    &16.420 &22.703 &--& $<18.00$ & GPS & V03 & P00 \\
 1615+028   &QSO & 1.339 &18.010 &17.750  &17.310 &15.890 &23.869&1 & $<18.30$ & FSSO &D85 & ---\\%agpw06
J1623+6624  &Gal &0.2030 &19.477 & ---    &17.430 &---    &20.004 &2 & $<18.48$ & HFP & O06 & ---\\
J1642+6856  &QSO &0.7510 &19.723 & ---    &19.219 &---    &22.667 &1 & $<18.10$ & EORG& V03 & M93\\
J1658+0741  &QSO &0.6210 &19.993 & ---    &19.598 &---    &22.441 &1 & $<18.17$ & EORG& V03 & M93\\
J1823+7938  &Gal &0.2240 &19.269 & ---    &17.415 &13.866 &20.385 &2 & $<19.44$ & GPS & V03 & T94 \\
J1829+4844  &QSO &0.6920 &16.260 & ---    &16.860 &14.250 &24.692 &1 & $<17.26$ & CSS & V03 & L98\\
J1831+2907  &Gal &0.8420 &21.917 & ---    &20.200 &---    &21.201 &2 & $<18.21$ & CSS & V03 & S89a\\
J1845+3541  &Gal &0.7640 &---    & ---    &21.900 &---    &---    &2 & $<19.02$ & GPS & V03 & X95 \\
  1937--101 &QSO &3.7870 &18.800 & ---    &17.188 &13.816 &24.910 &1 & $<18.11$ & --- & C08 & VLBA\\
J2022+6136  &Gal &0.2270 &19.830 & ---    &18.146 &---    &20.334 &2 & $<17.56$ & GPS & V03 & F00 \\
J2137--2042 &Gal &0.6350 &20.400 & ---    &19.286 &---    &21.808 &1 & $<18.05$ & CSS & V03 & T02,F03 \\
  2149+056  &QSO &0.7400 &23.700 & 22.050 &20.850 &17.170 &19.582 &1 & $<18.38$ & FSRS& C98 & S89b\\%J2151+0552
   2215+02  &QSO &3.5720 &21.840 & 20.420 &20.140 &19.340 &23.613 &1 & $<17.57$ &FSSO & C08& VLBA\\%agpw06
 J2250+1419  &QSO &0.2370 &16.760 & 16.640 &17.243 &---    &23.616 &1 & $<18.09$ & CSS & V03 & ---\\
 2300--189  &Gal &0.1290 &18.430 & ---    &16.569 &13.060 &20.099 &1 & $<18.20$ & --- & C06 & VLBA\\
J2321+2346  &Gal &0.2680 &20.315 & ---    &18.468 &14.710 &20.187 &--& $<18.28$ & EORG& V03 & G88\\ 
J2344+8226  &QSO &0.7350 &21.769 & ---    &20.220 &15.850 &21.165 &2 & $<17.85$ & GPS & V03 & D95,S01b \\
\tableline
\end{tabular}
\end{center}
\end{table*}
As stated above, at redshifts of $z\gapp3$ \citet{cww+08} detected no
21-cm absorption down to sensitivities sufficient for most of the
current detections, $N_{\rm HI}\lapp10^{18}.\,(T_{\rm s}/f)$ \scm\ per
channel (Tables \ref{dets} \& \ref{non-dets}, where we include the
$z\geq0.1$ searches which were previously
missed\footnote{\label{new_sources}We have added the $z\geq0.1$
  detections 1549--79 \& 3C\,433 and the $z\geq0.1$ non-detections
  0035--024, 0723--008, 1356+022 \& 1615+028, where the photometry and
  AGN type have been obtained/determined from
  \citet{awh82,wp85,lnn88,sh89,dwf+97,fww00,tdm+02,bmbd06,scs+06,aaa+08}.}).
From an analysis of the optical photometry, the target absorption
systems of \citet{cww+08} are found to be located in quasars with
high ultra-violet ($\lambda\approx1216$ \AA) luminosities ($L_{\rm
  UV}\gapp10^{23}$ \WpHz, Fig.\ref{lum-z})\footnote{Throughout this
  paper we use $H_{0}=71$~km~s$^{-1}$~Mpc$^{-1}$, $\Omega_{\rm
    matter}=0.27$ and $\Omega_{\Lambda}=0.73$.}, which suggests that
  the gas may have a significant ionisation fraction in all of these
  sources. In light of this, the large non-detection rate at high
  redshift is perhaps expected due to the flux limited nature of
  optical surveys, which selects only the most UV bright objects at
  these distances. Note, however, that this effect is also apparent
  for previously searched lower redshift ($z\lapp1$) sources, a trait
  which was previously unknown.

\begin{table*}
%\begin{minipage}{100mm}
\begin{center}
\caption{The incidence of detections for various UV luminosity
  partitions. 
\label{probs}}
\begin{tabular}{l  cccc  cccc} 
\tableline\tableline
$L_{\rm UV}$ & \multicolumn{4}{c}{For luminosities $<L_{\rm UV}$} & \multicolumn{4}{c}{For luminosities $>L_{\rm UV}$} \\
\WpHz & $k/n$ & rate & $P(\leq k/n)$ & $S(\leq k/n)$ & $k/n$ & rate & $P(\leq k/n)$ & $S(\leq k/n)$ \\
\tableline
% 10 or less out of 16, etc.
$10^{20}$ & 10/17 & 59\% & 0.31 &$1.00\sigma$ & 21/68 & 31\% &0.0011 & $3.26\sigma$\\
$10^{21}$ & 21/43 & 49\% & 0.50 & $0.67\sigma$& 10/41  & 24\% &0.00073  & $3.38\sigma$\\
$10^{22}$ & 25/53 & 47\% & 0.39 & $0.86\sigma$& 6/31  & 19\% &0.00044  & $3.52\sigma$\\
$10^{23}$ & 31/67 & 46\% & 0.31 & $1.00\sigma$& 0/17  & 0\% & $7.6\times10^{-6}$ & $4.46\sigma$\\
%%%%%%%%%%%%%%% new instructions for getting numbers %%%%%%%%%%%%%
%Log luminosity cut-off for statistical tests [e.g. 21.5]? 23
%emacs low-lum.txt &       DETECTION/NON CHANGE AT 
%                          J2355+4950  z = 0.238 logUV = 18.940, 2
%                          J0003+2129  z = 0.452 logUV = 20.971, 0
%
%%%%%%%%%%%%%%%%%%%%%%%%%%%%%%%%%%%%%%%%%%%%%%%%%%%%%%%%%%%%%%%%%%%%%%%%%%%%%%%%%
\tableline
\end{tabular}
\tablecomments{$k$ is the number of 21-cm detections below/above the
  partition and $n$ is the total number of sources in the same region,
  ``rate'' expresses this as a percentage detection rate, $P(\leq
  k/n)$ is the binomial probability of this number of detections or
  less occuring by chance for an unbiased sample and $S(\leq k/n)$ is
  the significance of this (derived assuming Gaussian statistics).}
%\end{minipage}
\end{center}
\end{table*}
In order to determine the significance of this UV segregation, in Table
\ref{probs} we summarise the binomial probabilities of the observed
distributions occuring by chance, given that a 21-cm detection and
non-detection are equally probable at a given UV luminosity. From
this, we see that below each UV luminosity cut-off there is no bias, with
the likelihood of a detection staying close to 50\%.
On the other hand, above the cut-off the probabilities of the observed
distributions resulting from an unbiased sample are small, dropping
dramatically at $L_{\rm UV}>10^{23}$ \WpHz. Since all luminosities
above the cut-off are included, the most luminous sources could well
dominate the upper partitions in Table \ref{probs}, which does appear
to be evident from the vertical histogram of Fig. \ref{lum-z}.  As is
also illustrated by the histogram, above $L_{\rm UV}\gapp10^{20}$
\WpHz\ the number of non-detections outweighs the number of detections
in each bin, which may suggest that at all values the ultra-violet
luminosity introduces a bias against 21-cm absorption. However, this
could also be explained by other effects which could make
non-detections more numerous, e.g. a larger proportion of type-1
objects (presuming that unified schemes were key in determining
whether 21-cm absorption could be detected, see below).  What is
clear, however, is that 21-cm has yet to be detected at $L_{\rm
  UV}>10^{23}$ \WpHz\ and that the probability of this distribution
occuring by chance is very small (Table~\ref{probs}).

\begin{figure*}
\centering \includegraphics[angle=0,scale=0.80]{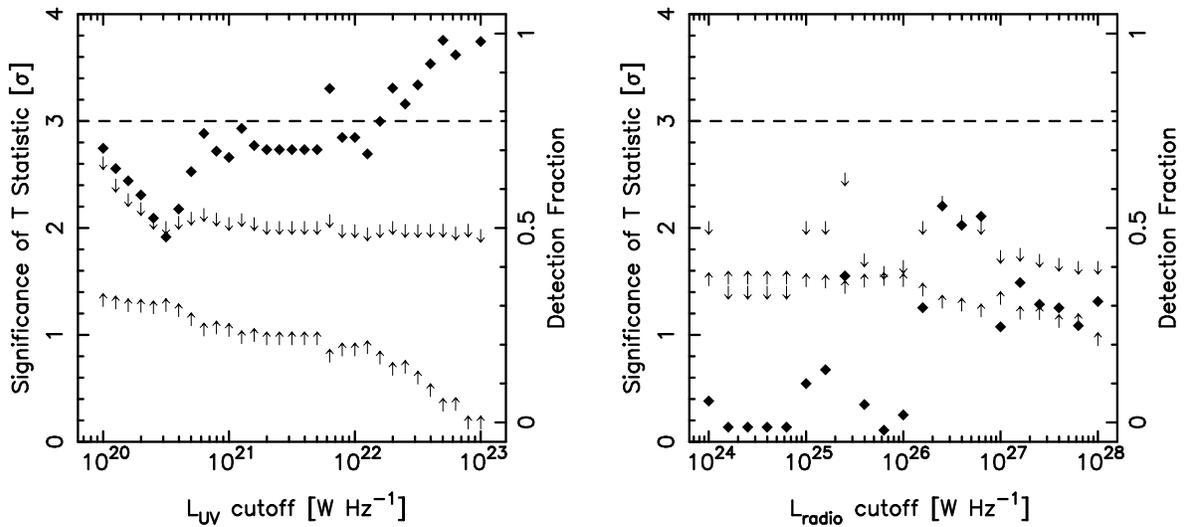}
\caption{The significance of the T-statistic of the difference in the
  proportions, $\hat{p}_1$ and $\hat{p}_2$, for various critical
  luminosities.  The left panel shows this for the ultra-violet and
  the right panel for the rest-frame 1420 MHz continuum
  luminosity. The ordinate label on the right-hand side of each panel
  shows the detection fraction above (upward arrows) and below
  (downward) the cut-off.}
\label{null}
\end{figure*}

A quantitative estimate of the critical UV luminosity may be obtained by
examining the detection proportions above and below certain values of
$L_{\rm UV}$. We define the statistic
%\[ 
\begin{equation}
T = \frac{\hat{p}_1 - \hat{p}_2}{\sqrt{\hat{p}(1-\hat{p})(N_1^{-1}+N_2^{-1})}},
\label{MrT}
\end{equation}
%\]
%$T = (\hat{p}_1 - \hat{p}_2)/\sqrt{\hat{p}(1-\hat{p})(N_1^{-1}+N_2^{-1})},$
where $\hat{p}_1=X_1/N_1$ and $\hat{p} _2=X_2/N_2$ are the two
measured proportions and $\hat{p}=(X_1+X_2)/(N_1+N_2)$ is the total
proportion. This has the standard normal distribution under the null hypothesis
that the proportions ($\hat{p}_1$ and $\hat{p}_2$) are the same, i.e.
that the UV luminosity does not affect the 21-cm detection rate. 
Testing the statistic in steps of $\Delta\log_{10}L_{\rm UV}=0.1$,
we reject the null hypothesis for all UV luminosities
of $\log_{10}L_{\rm UV}\geq22.5$, where the
difference between the two proportions is
significant at $\geq3\sigma$ (Fig. \ref{null}, left).
Additionally, as found above, the sources below the $L_{\rm UV}$
cut-off always exhibit a $\approx50$\% detection rate (Table
\ref{probs}), which is the natural rate in the absence of high UV
luminosities.

\subsubsection{Radio luminosity}
\label{rl}

Since, in the optically thin regime, the column density of the neutral
gas is related to the 21-cm absorption strength via $N_{\rm HI}\propto
(T_{\rm s}/f).\int \tau dv$, where $T_{\rm s}$ is the spin temperature
of the absorbing gas and $f$ the covering factor of the background
flux, a possible cause of the non-detections at high redshift could be
elevated spin temperatures, a trait which may have been observed in
intervening absorption systems \citep{kc02}.  However, for these, the
mean value at $z_{\rm abs} \gapp1$ may only be double that at
$z_{\rm abs} \lapp1$ ($T_{\rm spin}/{f}=2400$~K, cf. $1200$~K,
\citealt{ctd+09}), a factor which can be accounted for by the
different line-of-sight geometry effects, introduced by a flat
expanding Universe \citep{cw06}. This could negate the need for
significantly higher spin temperatures in order to explain the lower incidence of
detections at high redshift.

\begin{figure}[hbt]
\centering \includegraphics[angle=270,scale=0.75]{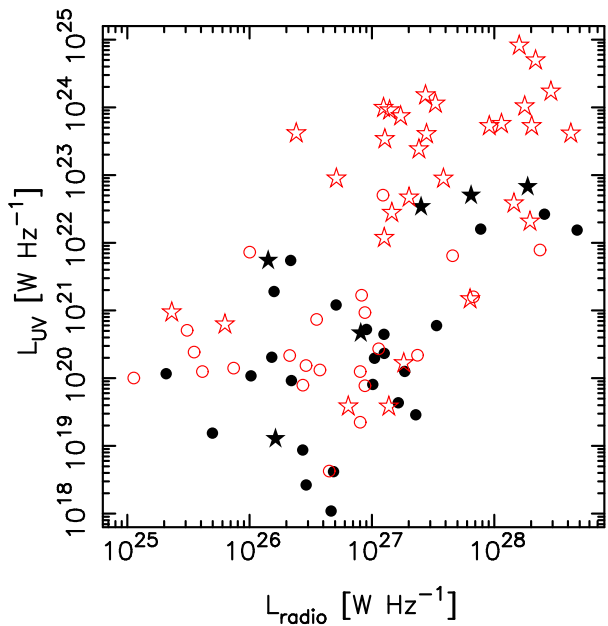}
\caption{The ultra-violet versus the radio luminosity for the sample. Again, the filled symbols
represent the 21-cm detections and the unfilled symbols the non-detections, with
stars signifying quasars and circles galaxies (see Sect. \ref{type}).}
\label{UV-radio}
\end{figure}

In the case of associated absorption, with $z_{\rm abs}\approx z_{\rm
  em}$ the same geometry effects cannot be responsible for such a
discrepancy between the low and high redshift samples, although the
generally higher radio luminosities of the latter could be raising the
spin temperature through a high population of the upper hyperfine
level.  We have already suggested that the more intense ultra-violet
fluxes could be responsible for the deficit in detections at high
redshift (Sect. \ref{lum}), which could be causing large ionisation
fractions and/or a raising of the spin temperature \citep{fie59,be69}
and so in Fig. \ref{UV-radio} we show the ultra-violet luminosity
(source-frame $\lambda\approx1216$~\AA) versus that of
the radio (source-frame $\lambda\approx21$~cm).

There is a strong apparent correlation between the two luminosities
($5.27\sigma$), with a large dispersion, particularly in the UV
luminosity. Any relationship will, however, be largely driven by both
quantities having a strong redshift dependence. However, we are not
sensitive to sources in the bottom right corner, where the optical
flux is too low to obtain a redshift (all sources in this sample are
identified as having associated absorption). Also, while all of the
21-cm detections are limited to $L_{\rm UV}\lapp10^{23}$ \WpHz, they
do cover the same range of radio luminosities as the non-detections,
thus suggesting that these are not so critical in the detection of
21-cm absorption. As a check, we apply Equ. \ref{MrT} to the radio
luminosities and find no significant relationship between the
detection rate and the radio continuum luminosity (Fig. \ref{null},
right). This is in contrast to the same statistic applied to the
ultra-violet luminosities and so it appears unlikely that the radio
power is the dominant effect in raising the spin temperature of the
gas above the detection thresholds.

%\clearpage
\subsection{Source structure and environment}

\subsubsection{Host galaxy environments}
\label{hge}

In considering the presence or otherwise of absorbing neutral gas, one
needs to consider the wider picture of the quasar host galaxy and its
environment. A great deal is known about the host galaxies of luminous
AGN at low redshift. Imaging studies, particularly with HST, of
$z\lapp0.4$ quasars \citep{tdhr96,bkss97,dmk+03,fkd+04},
indicate that early-type or spheroidal hosts are much more likely for
high-luminosity quasars, and that almost all luminous radio-loud
quasars and radio galaxies are to be found in elliptical galaxies.

Studies at higher redshift, often with ground-based adaptive optics
\citep{hcm+99,kdm+01,rhcl01,pir+06} show a slightly more complex
picture. The luminous quasars tend to follow the same trend, with
elliptical/early-type hosts predominant, although it is more difficult
to gain a confident picture of the host morphology at these redshifts.
Using gravitationally-lensed quasars can provide some extra
resolution, from which \citet{pir+06} find that quasar
hosts at $1<z<4.5$ span a range of morphologies from early-type to
disky/late-type galaxies.

Determining the expected \HI\ content of typical quasar host galaxies
is difficult, as 21-cm absorption is often the only way to probe the
neutral gas at such redshifts. For low redshifts, 21-cm emission
studies are possible with much being gleaned from blind surveys: The
HIPASS survey had detection rates of 6\% for RC3 ellipticals and 13\%
for S0s \citep{som02}\footnote{Although 30--50\% of these were
  confused, with more than one optical galaxy in the beam.} and from
the ALFALFA survey, \citet{dgg+07} detected \HI\ in just 2-3\% of
bright early-type galaxies in the Virgo cluster, with only one of
these (M86, an S0) having $M_{\rm B}<-20$ ($L_{\rm
  B}\gapp2\times10^{22}$ \WpHz). \citet{gdg+09}, however, examined
early-type galaxies in low-density environments, and found a higher
detection fraction of \HI\ emission for the more luminous objects,
although neither of the two ellipticals with $M_{\rm B}<-20$ were
detected.

Targeted searches with interferometers
\citep{oms+02,oms+07,mdo+06,nyl09} show different results. The
detection fraction tends to be much larger than in the blind surveys,
with a mix of ordered, relaxed disks and more complex interacting
systems seen. However, these have mostly been limited to fairly local
galaxies, although a small number of more powerful (yet still nearby
compared to our sample, $z<0.03$) radio galaxies have been observed to
have \HI\ disks as well as tails, resulting from an interaction, which
often coincide with \HI\ absorption. One such example is NGC612
\citep{emo+08}, an S0 galaxy with a large neutral disk of gas, likely
originating in an interaction or merger. However, hosts luminous AGN
are very rare in the local universe, making a clear understanding of
the \HI\ properties elusive.

The above can provide qualitative description of why we see the
cut-off in the absorption distribution at high luminosities, where the
the hosts tend to be the larger elliptical galaxies. For
instance, the low-redshift luminous AGN imaged by the HST all have
elliptical hosts \citep{bkss97,dmk+03,fkd+04} and $M_{\rm V}\lapp-23$,
which corresponds to a $V$-band luminosity of $L_{\rm V}\gapp10^{22}$
\WpHz\ -- similar to the brightest sources in our sample. Regarding
the $L_{\rm V}\gapp10^{23}$ \WpHz\ sources of our sample, of the five
imaged, two have resolved host galaxies both of which are ellipticals
(J1540+1447; \citealt{uso+00} and J2250+1419;
\citealt{dmk+03})\footnote{Of the remaining three, J0201-1132 and
  J1829+4844 are only marginally resolved (\citealt{kf00,lms+99},
  respectively) and in J0927+3902 no host is seen \citep{csg+98a}.}. 

To summarise, at lower luminosities the observed morphological mix
could present a range of \HI\ column densities, whereas at higher
luminosities the hosts are expected to be gas-poor ellipticals,
resulting in a lack of 21-cm absorption. This could give a mix
of 21-cm detections at low redshift and exclusive non-detections
at high redshift (Fig. \ref{lum-z}).
However, the bulk of the
$z\geq0.1$ sample has not been observed with a resolution sufficient
to determine the host galaxy morphologies, so that, while we do have
a good understanding of how the 21-cm detections are distributed with
$L_{\rm UV}$, how these are distributed with morphology for this
sample is at present undetermined.

%%%%Even when there are interactions present (with scenarios
%%%%such as NGC612), the distribution of \HI\ will be highly asymmetric,
%%%%and so the probability of getting a large covering factor will be
%%%%low. Whereas at the lower luminosities, it is
%%%%likely that the observed morphological mix presents a range of
%%%%\HI\ column densities, giving the observed mix of 
%%%%detections and non-detections at low redshift (Fig. \ref{lum-z}).

%\newpage
%\clearpage
\subsubsection{Radio source sizes}
\label{rss}

In addition to any possible evolution in the host galaxy environments,
selection effects may arise from the large range of beam sizes and
radio source sizes over what constitutes a heterogeneous sample (all
radio sources at $z\geq0.1$ searched for 21-cm absorption). For
example, a $10''$ diameter beam (a typical lower value for the GMRT at
90-cm, \citealt{cww+08}) covers a linear extent of 18 kpc at $z=0.1$,
$\approx86$ kpc (the maximum) at $z=1.6$ and 63 kpc at $z=5.2$. That
is, the area subtended by a given telescope beam can vary by a factor
of $\approx20$ due to the redshift range alone, an effect which is
compounded by the published searches being performed with numerous
instruments/configurations, each with its own beam size.  This can, in
principle, affect the sensitivity to 21-cm absorption. For instance,
absorption can be missed in near-by Seyfert galaxies by lower
resolution observations, while being revealed on VLBA/MERLIN scales
(e.g. \citealt{mwpg03}).
%%%%%\footnote{It is also possible systematic increase 
%%%%%in the spin temperature of the gas with redshift \citep{kc02} could
%%%%%also contribute to the observed deficit in 21-cm at high redshift,
%%%%%although as noted in Sect. \ref{rl}, $(T_{\rm
%%%%%  s}/f)_{z\geq1}\approx2\,(T_{\rm s}/f)_{z<1}$, which is consistent
%%%%%with geometry effects decreasing the covering factor with angular
%%%%%diameter distance. Furthermore, although we have ruled out that there
%%%%%is a correlation between the radio luminosity and the non-detection of
%%%%%21-cm absorption, raising of the spin temperature is possible through
%%%%%larger ultra-violet fluxes observed at high redshift
%%%%%\citep{fie59,be69}.}.

Therefore, the radio source sizes and the bias these may have on the
effective coverage of the background emission must be considered.
This, however, is fraught with uncertainties as the heterogeneity of
the sample combined with the very different scales probed, will
contribute to diluting out any strong trends in the linear extents of
the sources. Further compounding this is the fact that the radio
emission can exhibit very different structure at different frequencies
and, although the sizes used are from the nearest available
frequencies,VLBI and VLBA continuum observations are typically at
$\gapp2$ GHz, significantly higher than that of the redshifted 21-cm
line. The only way to fully address this would be through dedicated
mapping of the 21-cm absorption (e.g. \citealt{lbs00,mwpg03}), although the
available VLBA bands would only allow this at redshifts of $1.27 \leq
z \leq 1.38$ and $3.15 \leq z \leq 3.55$, thus covering only six of
the sample (all of which are non-detections in any case).

\begin{figure}
\centering \includegraphics[angle=270,scale=0.70]{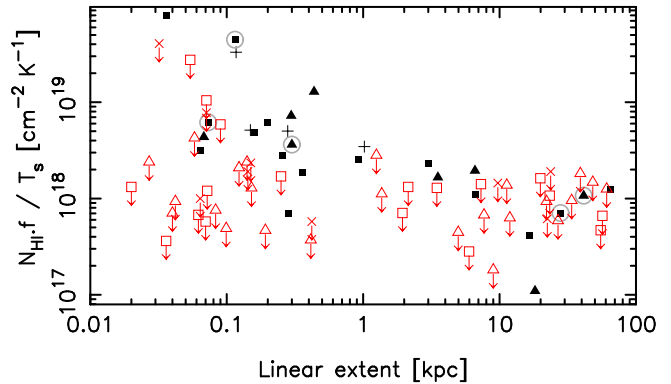}
\caption{The scaled velocity integrated optical depth
  ($1.823\times10^{18}.\int \tau dv$) of the 21-cm absorption versus
  the linear extent of the radio source size (from the references
  listed in Tables \ref{dets} and \ref{non-dets}.). The symbols are as
  per Fig. \ref{lum-z}, where the downwards arrows signify the upper
  limits to the non-detections. The five detections 
  not classified as compact objects (CSO, CSS, GPS nor HFP,
  defined in Sect. \ref{drco}) are circled.}
\label{tau-size}
\end{figure}
Nevertheless, applying these source sizes, in Fig. \ref{tau-size} we
see the same ``column density'' (actually the velocity integrated
optical depth)--linear size anti-correlation reported for the GPS and
CSS sources by \citet{pcv03,gss+06}: A Kendall's $\tau$-test on the
detections gives a two-sided probability $P(\tau) = 1.82\times10^{-5}$
of the correlation arising by chance, which is significant at
$4.28\sigma$ assuming Gaussian statistics. Including the 21-cm
non-detections, through the {\sc asurv} survival analysis package
\citep{ifn86}, decreases this to $P(\tau) = 0.017$ ($2.39\sigma$).  If
we consider just the GPS and CSS sources \citep{pcv03}\footnote{Although from Fig. \ref{tau-size} it appears that
{\em all} of the detections follow the same $\int \tau dv$--linear
  size correlation as is seen for the compact objects only
  \citep{pcv03,gss+06}, although this is based upon the small number
  of non-compact objects (five).}, we
obtain $P(\tau) = 0.006$ ($2.75\sigma$), which nonetheless indicates
that the inclusion of the non-detections significantly worsens the
correlation.  This suggests that the 21-cm non-detections may be
subject to an additional effect. This could be due to smaller
absorption cross-sections further reducing the covering factor,
generally higher spin temperatures and/or lower neutral hydrogen
column densities, all of which could be due to both orientation and UV
luminosity effects.

%\end{itemize} STUFF THAT WAS HERE IN rv-17nov.tex AND
%/home/sjc/papers/43/apj-paper.tex AND rv-dec09.tex

Although the decrease in the ``column density'' is most likely due to
a decrease in the covering factor with increasing with linear extent,
there is also the possibility that the 21-cm absorption towards the
larger sources is more susceptible to dilution by the extended 21-cm
emission. However, in spite of the four orders of magnitude span in
size, we find no significant difference in the detection rates between
the smaller ($<1$ kpc, 16 out of 42) and the larger ($\geq1$ kpc, 11
out of 36) sources (Fig.~\ref{tau-size}), indicating that this is not
a strong effect.

\subsection{AGN spectral type and source classification}
\label{type}

If the 21-cm non-detections are due to orientation effects, it would
suggest that all of the $L_{\rm UV}\gapp10^{23}$ \WpHz\ sources are
type-1 objects and therefore the gas may only necessarily be ionised
along our line-of-sight, which is direct to the AGN.  To examine this
question, we make use of the spectral classifications obtained by
\citet{cww+08}. These were compiled by exhaustively searching the
literature for published emission-line fluxes or spectra. Sources with
broad permitted lines were classified as type-1 and those with only
narrow lines as type-2. Our classifications generally agree with those
of \citet{vpt+03}, except for 2316+0405 which we assign as type-2
\citep{mil81}, although \citet{vpt+03} have this classified as a broad
line radio galaxy.

Using these classifications, it is
apparent that all the high UV luminosity sources are indeed type-1
(Fig.\ref{lum-z}).
However, they appear to be distinct from the
population of lower UV luminous type-1 objects, some of which have
been detected in 21-cm absorption at, somewhat surprisingly, the 
same detection rate (50\%) as for the type-2 objects. That is, the overall
bias towards type-1 non-detections is caused by the inclusion of the $L_{\rm
  UV}\gapp10^{23}$ \WpHz\ non-detections. Therefore, the ultra-violet
luminosity of the object appears to have much more bearing on whether
21-cm absorption can be detected and, at moderate UV luminosities, the
AGN type provides no indicator of whether a high column of cold
neutral gas is likely to intercept our sight-line.

Nevertheless, as has been noted in the literature
(e.g. \citealt{pcv03,gs06a}), 21-cm absorption is more likely to be
detected in a radio galaxy than in a quasar, even if there are even
odds between the two AGN types. We designate each object in our sample
as either a quasar or a galaxy, using the classifications from
\citet{cww+08}. To obtain these, optical imaging and spectroscopy from the
literature were used to distinguish sources dominated by the nuclear
source (the ``quasars'') from those dominated by the extended stellar
light of the host galaxy (the ``galaxies''), the latter being either
weaker AGN or relatively obscured type-2 AGN.

\begin{table}
\begin{center}
\caption{The mean ultra-violet luminosities [\WpHz] for the galaxy and
quasar sub-samples. $\sigma$ gives the standard deviation.
\label{uv-lum}}
\begin{tabular}{l cc cc } 
\tableline\tableline
 &\multicolumn{2}{c}{GALAXIES} & \multicolumn{2}{c}{QUASARS}\\
               & Detections & All & Detections & All \\
\tableline
$\overline{\log_{10} L_{\rm UV}}$ &  20.3 &  20.4  & 21.6 & 22.7 \\ 
$\sigma$ of $\log_{10} L_{\rm UV}$& 1.1 &   1.1  & 1.3  & 1.5  \\
Sample size&                      25 &   48    & 6   & 36    \\
%run sd on gal-all-lum.txt etc
\tableline
\end{tabular}
\end{center}
\end{table}
From Table~\ref{uv-lum}, where we show the average UV luminosities for
each class, we see that there is little difference in the luminosities
between the detections and whole sample for the galaxies (as well as a
50\% detection rate). However, the quasars in which 21-cm has been
detected are generally an order of magnitude brighter than the
galaxies, with the whole quasar sample being another order of
magnitude brighter yet (with only a 17\% detection rate). Again, over
and above an underlying 50\% due to orientation effects, this strongly
suggests that it is the ultra-violet luminosity which is the key
criterion in the detection of \HI\ in these objects and that any
detection bias against the quasars is due to their generally higher
ultra-violet output.

\subsection{Detection rates in compact objects}
\label{drco}

Until the 725--1200 MHz survey of \citet{vpt+03}, there were few
detections of associated 21-cm absorption at $z\geq0.1$. Of the
\citet{vpt+03} detections most (17 out of 19) are gigahertz peaked
spectrum and compact steep spectrum sources, with the general consensus
being that these exhibit higher 21-cm detection rates than other radio
sources, due to their gas rich-host galaxies
(\citealt{con96,ode98,gss+06} and references therein).  CSSs and GPSs
are believed to be intrinsically small ($\lapp10$ and $\lapp1$ kpc,
respectively) and may be the young precursors of the typically large
radio sources, themselves evolving from compact symmetric objects
(CSOs, \citealt{ffd+95}). Although the high occurrence of broad
forbidden lines may suggest that these are primarily type-1 objects,
in which the compact appearance is due to the radio lobes being
directed along our sight-line, the diminished sizes are not believed
to be due to projection effects \citep{ffs+90}: Although there are
radio lobes present, the jets may be embedded in a dusty interstellar
medium so that the AGN is believed to be subject to significant
extinction (e.g. \citealt{btm+03}), resulting in strong radio emission
as the trapped jets interact with the rich, dense cocoon, in this
early evolutionary stage.

The possibility that CSSs and GPSs are compact type-1 objects, would
be consistent with the 21-cm absorption being, on average, blue-shifted
with respect to the optical redshift (\citealt{vpt+03}, see
Sect. \ref{adto}). Furthermore, the efficient coverage of the confined
radio core could contribute to a high incidence of 21-cm absorption,
which probably occurs in an outflow. However, of the 17 CSS/GPS
detections of \citet{vpt+03} only six are classified as type-1
objects, cf. nine type-2s (and two unclassified, Table~\ref{dets}),
which again may suggest that these objects have random orientations,
with the possibility of absorption arising in either an outflow or the
disk.

If 21-cm absorption favours compact objects (and this was a more
important effect than the UV luminosity), we may expect a very low
number of CSSs/GPSs in the exclusively non-detected $L_{\rm
  UV}\gapp10^{23}$ \WpHz\ sample.  However, three of the eight low
redshift $L_{\rm UV}\gapp10^{23}$ \WpHz\ objects of \citet{vpt+03} are
classified as CSS/GPS, and so these are not immune to the 21-cm
absorption being undetected at high UV luminosity.  Being from the
Parkes Quarter-Jansky Flat-spectrum Sample, the high redshift sources
are, by definition, flat spectrum (with $\alpha>-0.5$), although of
these eight the radio SEDs for 1351--018 and 1535+004 are suggestive
of GPSs or high frequency peaker galaxies (HFPs)\footnote{The turnover
  frequencies of $\gapp5$ GHz may suggest newly born radio sources
  \citep{dal03} and from the anti-correlation between turnover
  frequency and the source size \citep{ffs+90,ob97}, we also expect
  these to be extremely compact. For the GPS/HFP suspects of the
  $L_{\rm UV}\gapp10^{23}$ \WpHz\ targets of \citet{cww+08}, the Very
  Large Array's FIRST (Faint Images of the Radio Sky at Twenty
  Centimetres, \citealt{wbhg97}) survey gives deconvolved source sizes
  of $<0.99''\times0.39''$ for 1351--018 and $<1.18''\times0.76''$ at
  an observed frequency of 1.4 GHz. At redshifts of $z=3.707$ and
  $3.497$, these sizes correspond to $<7\times3$ kpc and $<9\times6$
  kpc.  respectively, where at a turnover frequency of $\sim10$ GHz,
  we may expect a source size of $\lapp1$ kpc \citep{ob97}.}. For the
remainder of the $L_{\rm UV}\gapp10^{23}$ \WpHz\ sample, however,
being flat spectrum sources \citep{dsm85,jws+02,vpt+03} may {\em also}
suggest that many of these are oriented end-on with respect to us,
consistent with their type-1 status (Sect.~\ref{type}).

\begin{figure}
\centering \includegraphics[angle=270,scale=0.55]{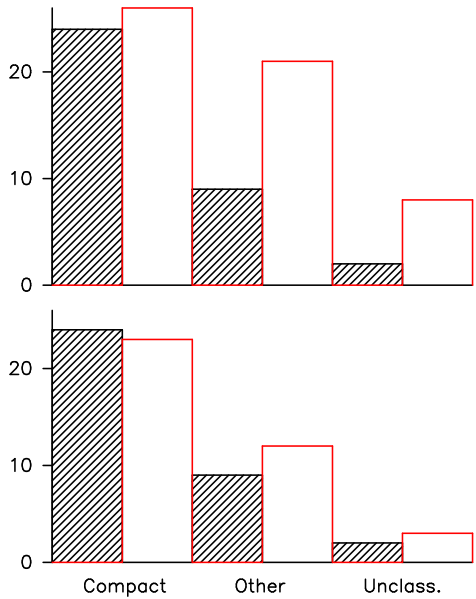}
% from compact-hist.c
% For all with L_uv > 0  24/26,  9/21, 2/8
%                        24/23,  9/12, 2/3    
% AND  z > 0.1           22/24,  7/21, 2/8  
%                        22/21,  7/12, 2/3
%%%%%%%%%%%%%%%%%%%%%%%%%%%%%%%%%%%%%%%%%%%%%%%%%%%%%%%%%%%%%%%%%%%%%
%\centering \includegraphics[angle=0,scale=0.45]{compact-hist-UV-2.eps}
\caption{The incidence of detections (hatched histogram) and
  non-detections (unfilled histogram) for compact objects (CSO, CSS, GPS
  and HFP) compared to the others. Top -- the whole sample. Bottom -- those with $L_{\rm UV}\leq10^{23}$ \WpHz.}
\label{compact-hist}
\end{figure}

Although the vast majority of the detections of \citet{vpt+03} are
CSS/GPS, 22 of the 38 non-detections are also classified as such
(giving a CSS/GPS detection rate of 44\%). Furthermore, \citet{pcv03}
and \citet{gs06a} both find a $\approx50\%$ detection rate in their
CSS/GPS samples, as well as the 33\% rate from the (albeit small) HFP sample of
\citet{omd06}. Summarising this in Fig. \ref{compact-hist} (top),
we find the overall detection rate to be 47\%, cf. 30\% for
the ``others'', not classified as CSO/CSS/GPS/HFP, and 20\% for those
unclassified, thus indicating higher detection rates in compact objects.

This distribution, however, becomes more uniform (51\% -- compact,
43\% -- others \& 40\% -- unclassified) between the classes when
considering the $L_{\rm UV}\leq10^{23}$ \WpHz\ sources only
(Fig. \ref{compact-hist}, bottom). That is, the compact source
detection rate {\em may not} be significantly higher than that of the others in the
moderate UV luminosity sample and, again, it is the inclusion of
$L_{\rm UV}\gapp10^{23}$ \WpHz\ sources which introduces a
bias\footnote{Although a larger number of the non-compact objects have
  luminosities of $L_{\rm UV}\leq10^{23}$ \WpHz, note that there are
  several CSSs/GPSs close to the UV cut-off (Table \ref{non-dets}).}.
\begin{table}
\begin{center}
\caption{The mean ultra-violet luminosities [\WpHz] and galaxy/quasar
distribution for the sample based upon radio classification.
\label{radio-class}}
\begin{tabular}{l c c c } 
\tableline
               & Compact & Other & Unclassified \\
    \tableline
$\overline{\log_{10} L_{\rm UV}}$ &  20.9 &  21.7  & 22.4\\
$\sigma$ of $\log_{10} L_{\rm UV}$& 1.4 &   1.7  & 2.0  \\
%$\overline{z_{\rm em}}$ & 0.60 & 0.67 & 2.5\\ 
No. galaxies&                      35 &   15    & 4 \\
No. quasars                     & 15  & 15  & 6 \\
Mean redshift $\pm1\sigma$& $0.60\pm0.96$ & $0.67\pm0.78$ & $2.5\pm1.4$\\ 
%FOLLOW INSTRUCTIONS WHEN RUNNING compact-hist.c
\tableline
\end{tabular}
\end{center}
\end{table}
This possibility is supported by the average UV luminosities of the
radio classes, which are appreciably lower for the compact objects
(Table \ref{radio-class}), which could be consistent with these being
young sources in which the AGN activity (radio and ultra-violet) has
yet to reach its full strength (Fig. \ref{UV-radio}).  Therefore, it
is possible that compact objects exhibit these higher 21-cm detection
rates due mainly to their generally low UV luminosities, with the line
strength in these being dominated by effect of the projection of the
radio lobes on the covering factor (Sect. \ref{rss}).

%%%%%Finally, although the samples of \citet{pcv03} and \citet{gs06a} were
%%%%%of moderate UV luminosity, giving their $\approx50$\% detection rates,
%%%%%both samples exhibit a preference for absorption in radio galaxies,
%%%%%cf. quasars, which considering the lower luminosities of the galaxies
%%%%%(Table~\ref{uv-lum}), may, again, indicate a UV luminosity effect.

\section{The location of the absorbing gas}

\subsection{\HI\ 21-cm line kinematics}
\subsubsection{Absorption in the galactic disk}

For the moderate UV luminous sample, which are not expected to be
dominated by elliptical hosts (Sect. \ref{hge}), the 50\% detection rate for {\em
  both} type-1 and type-2 objects at $L_{\rm UV}\lapp10^{23}$
\WpHz\ strongly suggests that the absorption {\em does not} occur in
the obscuring torus associated with the AGN.  The next logical
candidate is therefore the large-scale galactic disk, where most of
the gas would be expected to reside, as is seen in absorption studies
of low redshift starburst and Seyfert galaxies.  In particular,
\citet{gbo+99} find that the 21-cm line strength is correlated with
the galaxy inclination in $z\leq0.04$ Seyferts.  Furthermore, although
they suggest that the incidence of 21-cm absorption is broadly
consistent with unified schemes of AGN, \citet{mot+01} acknowledge
that it is likely that some absorption is also occuring beyond the
sub-pc scale of the circumnuclear torus. Although we may expect the
gas to be nearly coplanar on all scales (e.g. \citealt{ckb08}), if the
large-scale disk is the source of the absorption, the fact that both
types of AGN have the same odds of exhibiting 21-cm absorption would
suggest that the orientation of the large-scale disk with respect to
the small-scale obscuring torus is random.

\begin{figure*}
\centering \includegraphics[angle=270,scale=0.70]{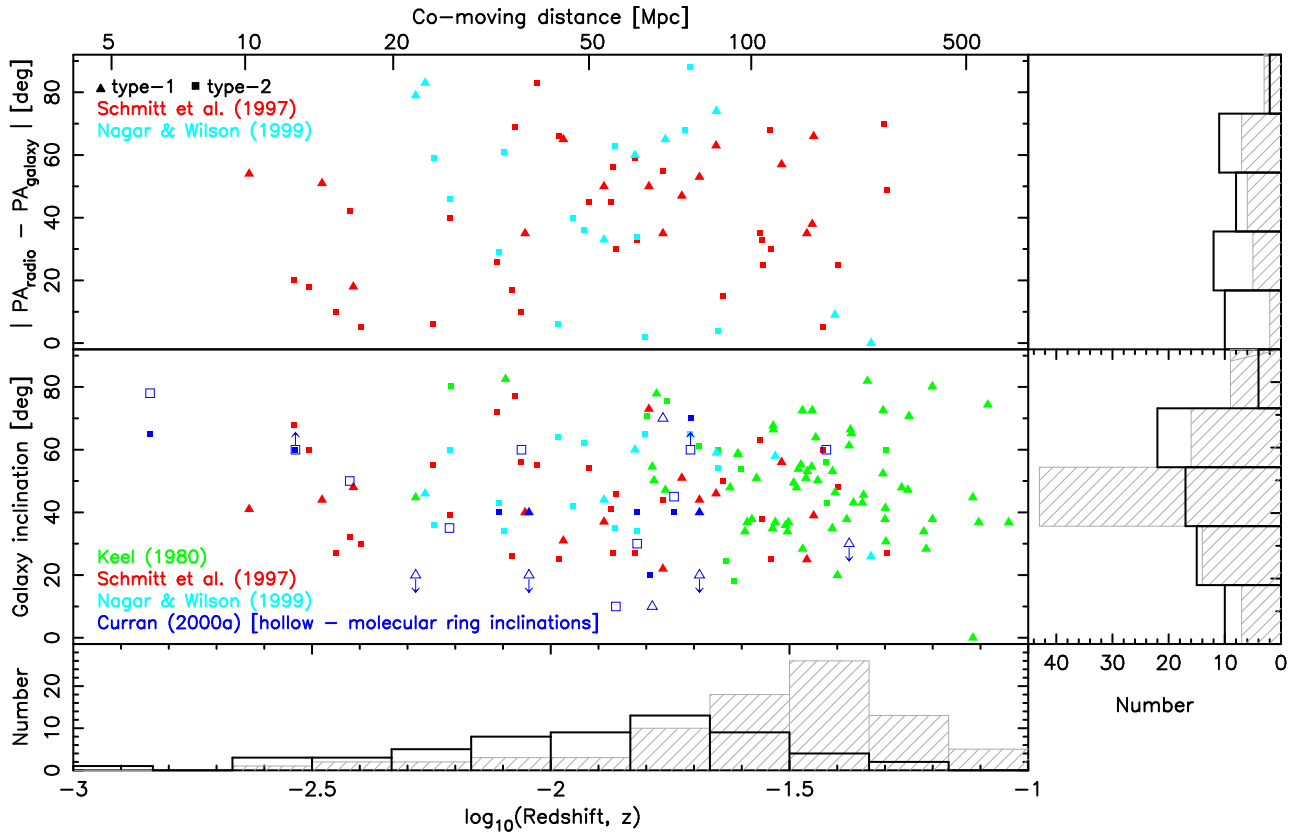}
\caption{The difference in the position angles of the radio and dis axes (top) 
and the inclinations of the galaxies (bottom) hosting low redshift
  Seyferts. The triangles/hatched histogram represent the type-1
  objects and the squares/unfilled histogram the type-2, with the
  colours of the symbols indicating the source reference
  \citep{kee80,sksa97,nw99,cur99p}.}
\label{inc-z}
\end{figure*}

At redshifts of $z\geq0.1$, the disk orientations cannot generally be
determined, although from lower redshift Seyfert studies we can show
the inclination of the galaxy disk for various Seyfert types
(Fig. \ref{inc-z}).
From the figure we see the full range of possible offsets between the
position angles of the radio jets and the position angle of the host
galaxy (top panel). Were the obscuring torus and main galaxy disk
coplanar, we would expect the points on the ordinate to be
concentrated close to 0\dg\ for both AGN types, but, as noted by
\citet{sksa97,nw99}, the distribution tends to be quite random.
Furthermore, if the torus is coplanar with the host disk, we would
expect type-1 Seyferts to occupy galaxies of low inclination and
type-2s in those of high inclination. However, from the vertical
histogram (Fig. \ref{inc-z}, bottom panel) it is apparent that neither
Seyfert type has a preferred disk inclination, with both types
exhibiting a mean inclination of 48\dg\ ($\sigma\approx16$\dg, from 83
exclusive\footnote{Sources common to more than one sample have
only been counted once.} type-1 objects and 58 exclusive
type-2s). Interestingly, although the large-scale disk exhibits a
random orientation with respect to the circumnuclear torus, the
sub-kpc molecular ring is generally aligned (the hollow symbols in
Fig. \ref{inc-z}): Although the numbers are much smaller, the mean
inclination of the molecular ring in the six type-1 objects is
$<28$\dg, cf. $>49$\dg\ for the ten type-2 objects
($\sigma\approx19$\dg, with the limits being due to limits in the
inclination estimates, \citealt{cur99p}). A Kolmogorov-Smirnov test
gives a $<6$\% probability that the inclinations of the type-1 and
type-2 molecular rings are drawn from the same sample, in contrast to
38\% for the galactic disk inclinations. This alignment between the
molecular ring and torus may be expected, despite the random larger
scale orientations, as these rings generally only reach $\sim1/100$th
the extent of the atomic gas
(\citealt{ms87,plr+91,babr92,is92,tgb+94,kkto96,cjrb98,ivha98})\footnote{Although
  the molecular gas beyond the ring can be seen to extend much farther
  (e.g. \citealt{ys91,ckb08}).} and may be funneling the material to
the smaller scale torus (see \citealt{cur99} and references
therein)\footnote{\label{foot7}http://nedwww.ipac.caltech.edu/level5/Curran/frames.html}.
%(e.g. \citealt{bs98}).
%REFS FROM THESIS

\subsubsection{Absorption due to outflows}
\label{adto}

Aside from the disk, as noted above, absorption may also be due to
in-falling gas or outflows. That is, gas located along the polar axes,
thus being located between us and the AGN in type-1 objects, rendering
the gas detectable in absorption. Since this gas {\em may} exhibit a
wider profile (FWHM) than gas tracing the rotation of the
disk\,\footnote{For example, presuming the systemic velocities are
  sufficiently accurate to determine these large blue-shifted offsets,
  outflows with widths of $\gapp1000$ \kms\ are seen in some low
  redshift radio galaxies \citep{mto05,mhs+07}, in addition to the
  broad components observed by \citet{moe+03,mot+05}, in which the
  blue-shifts from the peak absorption component are strongly
  suggestive of outflows.}, while having a larger offset from the
systemic velocity of the galaxy ($\Delta v$), we may expect type-1
objects to be grouped separately from type-2s in a plot of FWHM versus
$\Delta v$, at least in terms of the abscissa.

\begin{figure*}
\centering \includegraphics[angle=270,scale=0.70]{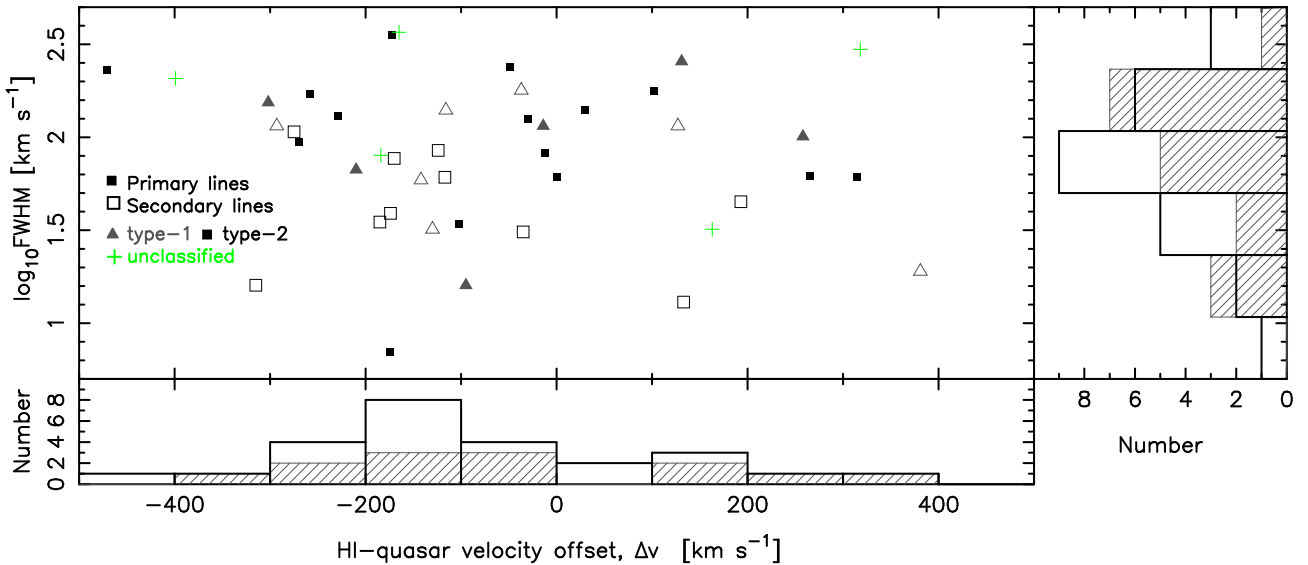}
\caption{The profile width versus the offset from the systemic
  velocity for the 21-cm detections at $z\geq0.1$. The
  triangles/hatched histogram represent the type-1 AGN and the
  squares/unfilled histogram the type-2. {\bf +} signifies that there
  is no AGN classification available. We have plotted each
  individually resolved absorption component and flagged the primary
  (filled symbols) and secondary (unfilled) absorption lines, where
  the primary is the line with the largest optical depth and the
  secondaries are the remaining shallower lines (see \citealt{vpt+03}).
  For the sake of clarity, the plot has been truncated to $|\Delta
  v|\leq500$ \kms, although there are five cases with $\Delta v<-500$
  \kms: 1413+135 at $-705$ \kms, J1815+6127 at $-1258$ \kms,
  J1821+3942 at $-869$ \kms\ (primary lines), as well as the $-742$
  \kms\ secondary line in the latter object. All of these lines arise
  in type-1 objects, with the one type-2 being in J1944+5448 with
  $-1420$ \kms\ (primary line). On the redshifted end there are just
  three cases at $v> 500$ \kms -- the two unclassified AGN J0431+2037 at $+636$
  \kms\ and 0902+343 at $+970$ \kms, as well as the type-1 object 1549--79 at $+665$
  \kms.}
\label{FWHM-detlaV}
\end{figure*}
Showing this in Fig. \ref{FWHM-detlaV}, we see no clear distinction
between the AGN types in either axis. For the FWHM, the type-1
absorbers do exhibit slightly wider profiles, which would suggest that
the outflows are subject to large velocity differentials, as has been
seen in several low redshift cases
(e.g. \citealt{moe+03,mot+05,mto05,mhs+07}). If absorption were
occuring in the circumnuclear torus, we may also expect
  very wide profiles for the type-2 AGN, possibly much wider than
  those of the galactic disk itself\,\footnote{For example, in the type-2
  Seyfert NGC 4258, the sub-parsec disk is found to have a rotation
  speed of $900$ \kms\ \citep{hbp94} and in the Circinus galaxy, also
  a type-2 Seyfert, the \WAT\ masers also trace a disk which rotates
  much more rapidly than the galactic disk (\citealt{gbe+03},
  cf. \citealt{ckb08}).}.
%, which if as deep as the disk absorption could completely mask this. 
As it is, we can make no distinction between the FWHM of the type-1
and type-2 objects and unlike in emission, the absorption profile
widths will ultimately be subject to the covering factor and the size
of the continuum source, making any distinction between disk and
outflow absorption difficult.

Again, for the velocity offsets there is no real difference between
the two AGN types and both exhibit a slight bias towards blue-shifted
absorption\footnote{This is most obvious for the type-2s in
  Fig. \ref{FWHM-detlaV}, but as stated in the caption, there are also
  four (three primary and one secondary) type-1 absorbers with $\Delta
  v<-500$ \kms.}. Although uncertainties of $\sim10^2$ \kms\ due to
the optical emission lines in $\Delta v$ are to be expected, many
studies show a tendency for the absorption to be blue-shifted with
respect to the systemic velocity
\citep{vpt+03,moe+03,mot+05,mto05,mhs+07}. If these offsets are
artifacts of poorly constrained optical redshifts, we would expect
similar numbers of red-shifted components and, as stated previously, in
low redshift radio galaxies wide blue-shifted tails are seen in the
profiles, where the main feature is close to the systemic velocity
(\citealt{moe+03,mot+05}).  

Since many of the 21-cm detections at $z\geq0.1$ are from
\citet{vpt+03} [Table~\ref{dets}], it is not surprising that we also
see this skew towards blue-shifted absorption
(Fig. \ref{FWHM-detlaV}).  
Therefore in Table~\ref{stats}, where we
show the average offsets for the AGN classes, we also show the
contribution from the remainder of the literature.  
\begin{table*}
\begin{center}
\caption{The means ($\overline{\Delta v}$) and standard deviations ($\sigma$) for the 
absorption offset from the systemic velocity [\kms].
\label{stats}}
\begin{tabular}{l ccc ccc ccc ccc ccc}
\tableline\tableline
 & \multicolumn{3}{c}{TYPE-1} & \multicolumn{3}{c}{TYPE-2} & \multicolumn{3}{c}{WHOLE SAMPLE} & \multicolumn{3}{c}{\citet{vpt+03}} & \multicolumn{3}{c}{OTHERS} \\
 & $\overline{\Delta v}$ & $\sigma$ & $n$& $\overline{\Delta v}$ & $\sigma$ & $n$& $\overline{\Delta v}$ & $\sigma$ & $n$& $\overline{\Delta v}$ & $\sigma$ & $n$ & $\overline{\Delta v}$ & $\sigma$ & $n$\\
\tableline
Primary   & -240 & 540 & 10 &  -150 & 380 & 16 & -130 & 470 & 32 & -220 & 470 & 19 & 10 & 420 & 13 \\
Secondary & -120 & 300 & 8  &  -110 & 150 & 10 & -70  & 280 & 19 & -60  & 380 &  9 & -90 & 150 & 10 \\
Both      & -190 & 460 & 18 &  -140 & 310 & 26 & -110 & 410 & 51 & -170 & 450 & 28 & -30 & 330 & 23\\
\tableline
\end{tabular}
\end{center}
\end{table*}

As seen from the table, although the standard deviations
are large, the additional results confirm that on average the
absorption is blue-shifted, which could indicate outflows or some
other non-symmetric mechanism as the origin. The statistics also
confirm the larger spread in the velocity offsets of the type-1 AGN,
which is not wholly evident from Fig.~\ref{FWHM-detlaV}, due to the
three (primary) type-1 absorbers offset at $\Delta v < -500$~\kms. If
it were just these three at the blue-shift extremes, we could at least
state that some type-1s show more of a bias for absorption in rapidly
outflowing gas, although there is the type-2 case (J1944+5448) with
$\Delta v = -1420$~\kms. This, however, could be the consequence of a
poorly constrained optical redshift, or a rapid outflow of gas located
well clear of the jet axis, as well as the possibility that this is
unassociated gas.

\subsection{Extinction effects}

\begin{figure}[hbt]
\centering \includegraphics[angle=0,scale=0.75]{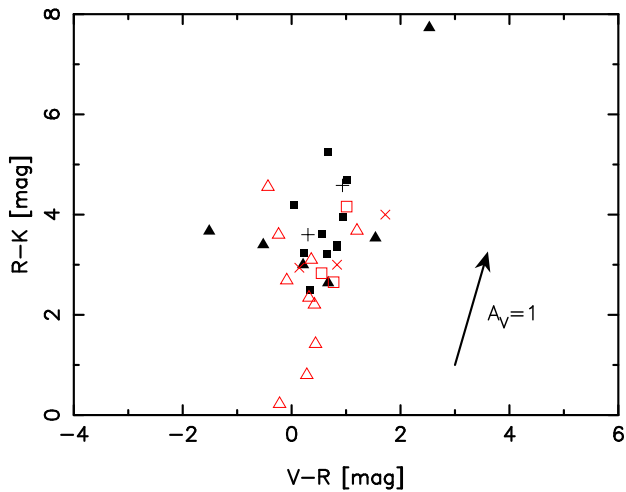}
\caption{The $R-K$ colour versus the $V-R$ colour for the sample. As
  per Fig. \ref{lum-z}, the filled symbols represent the 21-cm
  detections and the unfilled symbols the non-detections, with the
  triangles representing type-1 objects and squares type-2s ({\bf +}
  and {\sf x} designating a non-specific AGN type for a detection and
  non-detection, respectively).}
\label{colourcolour}
\end{figure}

\begin{figure*}
\centering \includegraphics[angle=0,scale=0.40]{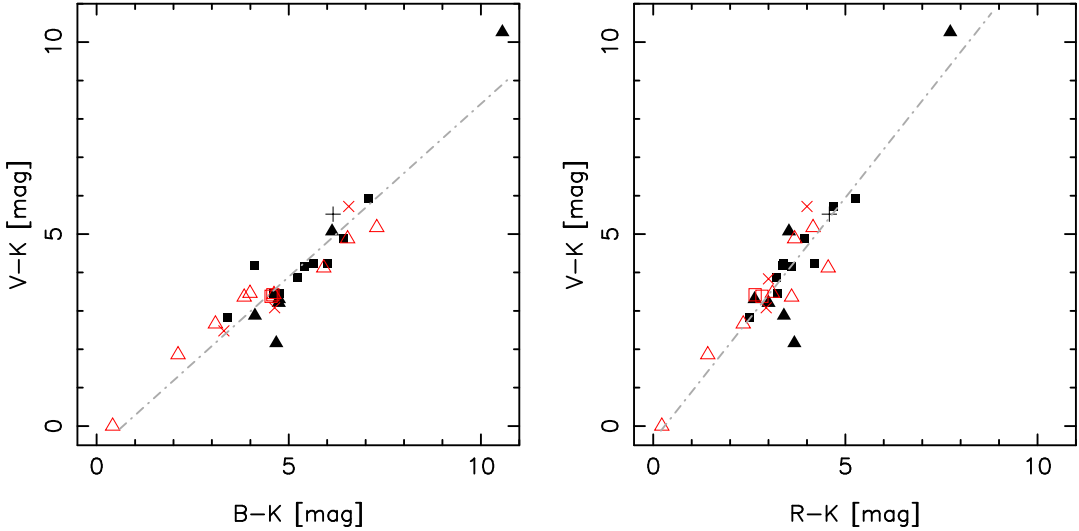}
\caption{Comparison of the $V-K$ colours with the $B-K$ and $R-K$
  colours for the $z_{\rm em} < 3$ sample (to avoid contamination of the $B$ and $V$ 
  bands by Lyman-\AL\ absorption). From the fits shown we obtain $V-K
  = 0.90\pm0.01\times(B-K) - 0.62\pm0.29$ (significant at
  $5.86\sigma$) and $V-K = 1.26\pm0.02\times(R-K) - 0.37\pm0.29$
  (significant at $5.25\sigma$), which are used to convert $B-K$ and
  $R-K$ to $V-K$, where $V$ is unavailable. The symbols are as per
  Fig.~\ref{lum-z}.}
\label{mag-comps}
\end{figure*}

\begin{figure*}
\centering \includegraphics[angle=270,scale=0.70]{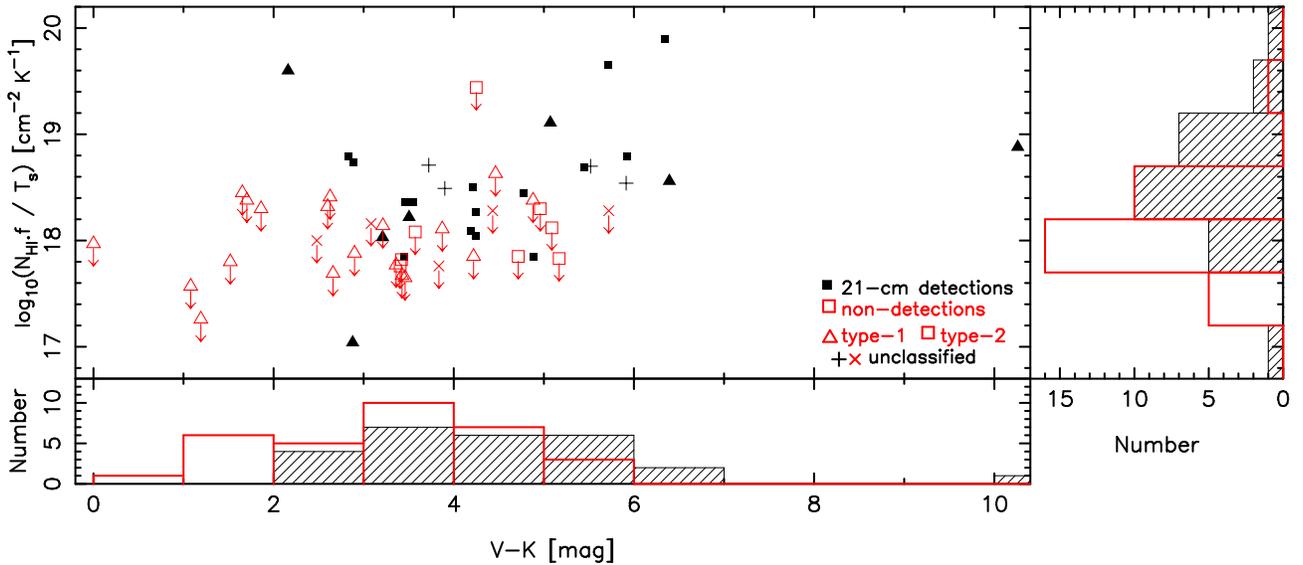}
\caption{The scaled velocity integrated optical depth of the \HI\ line
  ($1.823\times10^{18}.\int \tau dv$) versus optical--near-infrared
  colour for the sample. The symbols are as per Fig. \ref{lum-z}, with
  the hatched histogram represent the 21-cm detections and the
  unfilled histogram the non-detections. Where the $V$ magnitudes are not available
  (Tables 6 \& 7 of \citealt{cww+08}), we have estimated these
  according to the fits derived in Fig. \ref{mag-comps}. The statistics are
  summarised in Table \ref{red}.  The outlier at $V-K = 10.26$ is due
  to J0414$+$0534, which has an intervening gravitational lens, which
  may be responsible for some of the extreme reddening (see
  \citealt{cdbw07} and references therein).}
\label{N-colour}
\end{figure*}

So far, at least for the intermediate UV luminosity sample ($L_{\rm
  UV}\lapp10^{23}$ \WpHz), we have found no difference in the
incidence of 21-cm absorption between the two AGN types, which are
also indistinguishable through the absorption line profiles of the
detections. In order to determine whether there is a difference in the
extinction of the quasar light, in Fig. \ref{colourcolour} we show the
$V-R$ and $R-K$ colours (where available) for the published
$z\geq0.1$ searches as classified by AGN type, 
where, apart from four
type-1 outliers, we see no discernible difference between the two
types\footnote{The outlier with the high extinction in the direction of the 
reddening vector is the extremely red quasar
  J0414$+$0534 (see Figs.~\ref{mag-comps}~and~\ref{N-colour}).}. This
suggests that the circumnuclear torus does not introduce a measurable
degree of optical extinction, although contamination from the host
galaxy starlight could lessen any apparent reddening.

\begin{table}
\begin{center}
\caption{The statistics from Fig. \ref{N-colour}. 
   \label{red}}
\begin{tabular}{l c rc cc c }
\tableline\tableline
Sample & $n$& \multicolumn{2}{c}{$\log_{10}(N_{\rm HI}\,.f/T_s)$} & \multicolumn{2}{c}{$V-K$} & $S(\tau)$ \\
&    &  Mean & $\sigma$                            &  Mean & $\sigma$ &  \\
\tableline
Whole      & 58 & $<18.3$ & 0.5 & 3.9 & 1.6 & $3.63\sigma$ \\
Detections & 26 & $18.5$ & 0.6 & 4.6 & 1.6 & $2.03\sigma$ \\
Type-1 detections & 7 & 18.5 & 0.8 & 4.8 & 2.6 & $0.75\sigma$ \\ 
~~~~~non-detections & 19 & $<18.0$ & 0.4 & 2.7 & 1.3 & --\\
~~~~~~~exc. UV lum. & 6 & $<17.9$ & 0.3 & 3.6& 0.7 & -- \\
Type-2 detections & 15 & 18.6 & 0.6 &  4.4 & 1.0 & $1.19\sigma$ \\
~~~~~non-detections & 8 & $<18.2$ & 0.5 & 4.3 & 0.7 & --\\
% RUN mean TO GET VALUES AS WELL AS asurv FOR LAST COLUMN
\tableline
\end{tabular}
\end{center}
\end{table}
In order to further examine the reddening, after correcting for the
unavailable $V$ magnitudes (see Fig. \ref{mag-comps}), in Fig. \ref{N-colour}
we show the 21-cm line strengths and limits against the
optical--near-infrared colour of the source.
From the statistics (Table \ref{red}), aside from the incidence of the
mix of detections and non-detections (as shown by $n$)\footnote{The
  numbers give the impression that 21-cm absorption is more likely to
  arise in type-2 objects (cf. Fig. \ref{lum-z}).  However these are
  subject to the available magnitudes and many of the $K$ magnitudes
  are unavailable for the non-detections (Table \ref{non-dets}).}, the
only significant difference between the AGN types are the $V-K$ colours for the
non-detections: Although the type-1 detections are slightly redder
than the type-2 ($V-K\approx4.8$ cf. $4.4$), the type-1 non-detections
could be considerably less red than those of the type-2s ($V-K = 2.6$,
cf. $4.3$). Naturally, this will be skewed by the inclusion of the
$L_{\rm UV}\gapp10^{23}$ \WpHz\ sources (Sect.~\ref{lum}) and omitting
these raises this to $V-K = 3.6$ (the ``exc. UV lum.''  row), which,
considering the $1\sigma$ spreads, is indistinguishable from the
type-2 values. Note finally, that a mean line strength of
$\log_{10}(N_{\rm HI}\,.f/T_s)\sim18.5$ is exhibited for the
detections of both AGN types.

On this issue, over the whole sample we find a correlation between the
21-cm line strength and the $V-K$ colour, significant at $3.63\sigma$,
Table \ref{red})\footnote{The upper limits in the 21-cm line strengths
  (final column) are incorporated via the {\sc asurv} package.}. This
  may suggest that the reddening is due to dust associated with the
  intervening neutral gas, rather than intrinsic to the AGN spectrum
  (\citealt{wfp+95,srkr96}; although see
  \citealt{fww00,wwf01}). Circumstantial evidence for this association
  was previously presented by \citet{cmr+98}, although we show, for
  the first time, a correlation between the line strength and
  colour. However, in light of our other findings, we know that the
  high UV luminous sources have low 21-cm line strengths, and these
  being located at the blue end of Fig.~\ref{N-colour}\footnote{All
    but two of the $L_{\rm UV}\lapp10^{23}$ \WpHz\ sources have
    $V-K<3.5$.} must drive much of the correlation. Furthermore, the
  overall correlation is quite fragile with the significance dropping
  quickly with the sample size (Table \ref{red}). This could be a
  reflection of the heterogeneous nature of the sample\footnote{And is
    thus not apparent in Fig. \ref{colourcolour}.}, some of which will
  also be subject to contamination from starlight, although we have
  previously found that the molecular, rather than atomic, gas content
  appears to dominate the degree of reddening \citep{cwm+06}.

\section{Conclusions and Interpretation}
\subsection{Ultra-violet luminosities}

In a previous paper \citep{cww+08} we found that the ultra-violet
luminosity plays more of a r\^{o}le than the AGN type in the detection
of 21-cm absorption in $z\geq0.1$ radio galaxies {\em and} quasars.
Specifically, to date, 21-cm absorption has never been detected in a
host galaxy when $L_{\rm UV}\gapp10^{23}$ \WpHz. Although these high
UV luminosities occur exclusively in type-1 objects, for the moderately
UV luminous ($L_{\rm UV}\lapp10^{23}$ \WpHz) sample there is a 50\%
probability of detection in {\em either} AGN type, with any apparent
bias against type-1 objects being caused by the 17 high UV
luminosity objects. Expanding upon these results, in this paper we
show:

\begin{enumerate}
  \item That the bias against 21-cm absorption in quasars, compared to
    radio galaxies, also appears to be due to UV luminosity effects:
    For the radio galaxies, the 21-cm detections and non-detections
    both arise in objects with a mean $\overline{L_{\rm
        UV}}\approx2\times10^{20}$ \WpHz, whereas the 21-cm detected
    quasars have $\overline{L_{\rm UV}}\approx4\times10^{21}$ \WpHz,
    with the non-detected quasars having $\overline{L_{\rm
        UV}}\approx5\times10^{22}$ \WpHz.

    \item Although there is this decrease in the 21-cm detection rate
      with increasing $L_{\rm UV}$, it is possible that the exclusive
      21-cm non-detections at $L_{\rm UV}\gapp10^{23}$ \WpHz\ are due
      to the fact that highly luminous sources are believed to trace
      the (neutral) gas-poor elliptical galaxies. However only two of
      the 17 $L_{\rm UV}\gapp10^{23}$ \WpHz\ sources are {\em known}
      to be located in ellipticals (although all of them could be),
      whereas all are known to have high ultra-violet luminosities.
      Therefore it is not clear whether the lack of 21-cm absorption is due
      to a low abundance of neutral galaxy in the host or excitation
      effects caused by the high luminosities, or indeed how these two
      scenarios may be related.

      \item With the exclusion of the $L_{\rm UV}\gapp10^{23}$ \WpHz\
        sources, the skew towards the detection of
        21-cm absorption in compact objects (CSO/CSS/GPS/HFP) 
        becomes much less significant. Again, this indicates a bias introduced by
        the high UV luminosity sample and perhaps suggests that it
        is more meaningful to discuss the 21-cm absorption incidence
        in terms of the rest frame ultra-violet luminosities rather than
        by AGN type or radio SED classifications.        

\end{enumerate}

\subsection{Detection rates at moderate UV luminosities}

%\begin{itemize}

Regarding the $L_{\rm UV}\lapp10^{23}$ \WpHz\ sources, in addition to
both AGN types having a 50\% detection rate, there is no evidence for
the expected larger degree of reddening in the type-2 objects, which
would be caused by the presence of a dusty obscuration. We do find,
however, that the optical--near-infrared colour appears to be
correlated with the 21-cm absorption line strength over the {\em
  whole} sample.
These points  would therefore appear to contradict
      the notion that the 21-cm optical depth in the hosts of high
      redshift galaxies and quasars is determined by the orientation
      of the central obscuring torus.  Although, like the literature,
      we find a higher incidence of 21-cm absorption in galaxies than
      in quasars, unlike the literature, we believe this to be a
      consequence of their lower ultra-violet luminosities, rather
      than their AGN classification\footnote{As discussed in
        \citet{cww+08}, the exclusivity of type-1 objects at $L_{\rm
          UV}\gapp10^{23}$ \WpHz\ is not likely to be coincidental.
        They do, however, seem to be quite distinct from their lower
        UV luminosity counterparts, which exhibit the same 21-cm
        detection rate as the type-2 objects.}. Furthermore, the fact
      that the reddening is also independent of AGN type, suggests
      that this is also unrelated to the torus and the correlation
      between the reddening and the absorption line strength suggests
      that the bulk \HI\ and dust share a similar (non-nuclear)
      distribution.  

      This could explain why only a fraction of
      \HI\ absorbing AGN are detection in \WAT-maser emission
      \citep{tph+02}: \WAT-masers are believed to arise close to the
      black hole in the central obscuration \citep{hbp94} and thus
      trace type-2 objects.  Therefore, if \HI\ absorption also arose
      in the torus, one would expect a high detection rate of
      \WAT-masers in AGN detected in \HI\ absorption. However,
      \citet{tph+02} find \HI\ and \WAT\ common to only 8 out of 19
      objects searched and this 42\% rate is very close to the overall
      \HI\ detection rate in AGN (see \citealt{cww+08}). This therefore
      suggests that the lines-of-sight through the masing disk and the
      \HI\ absorbing clouds are randomly oriented with respect to one another.

In the moderate UV luminous sample, not dominated by elliptical hosts,
we therefore argue that the absorption may be occuring in the main
galactic disk which must be randomly oriented with respect to the
torus.  In attempting to verify this:
      \begin{enumerate}
      \item We find no real difference in the full-width half maxima
        of the 21-cm absorption profiles between the two AGN types.
        This may suggest that these are subject to geometric effects
        (covering factors and radio source sizes) and thus cannot give
        the full kinematical picture. Although, again, absorption due to 
        a randomly oriented disk could account for this.

        \item We also find no discernible difference in the offset of
          the centroid of the absorption and the systemic velocity of
          the host between the two AGN types. Although through a
          sample nearly double in size, we can confirm the findings of
          \citet{vpt+03}, that on average the offsets are blue-shifted.
          That is, there {\em may be} evidence for outflowing gas,
          although, again, we see no distinction between the two AGN types.
 \end{enumerate}        
%\end{enumerate}
If the galactic disk were aligned with the obscuring torus, the 50\%
detection rate for {\em both} AGN types suggests that absorption must occur
in both galactic disks {\em and} outflows: The outflowing gas is
expected to be directed along the radio jets
(e.g. \citealt{bbr84,sch88,cbg+96,cbov98}), which are coincident with
the axis of the torus, thus giving rise to the absorption in the
type-1 objects. The fact that we cannot discriminate between the disk
and outflow absorption would suggest rapid, wide outflows of cold
neutral gas, perhaps enveloping the wide ionisation cones observed in
low redshift AGN (see table 1.2 of \citealt{cur99})$^{\ref{foot7}}$.
A 90\deg\ wide outflow of cool neutral gas expanding at $\approx200$
\kms, in which the molecular gas mass rivals that in the disk ($M_{\rm
  H_2} \sim10^{9}$\Mo), is known of in the Circinus galaxy
\citep{crjb98}, the proximate type-2 Seyfert (see also
\citealt{moe+03,mot+05} for further examples).

However, it would therefore remain a mystery as to why only 50\% of
each AGN type (of $L_{\rm UV}\lapp10^{23}$ \WpHz) are detected in
21-cm absorption, although the bulk absorption occuring in the
galactic disk, which is randomly oriented with respect to the
obscuring torus, could account for this.\\

This research has made use of the NASA/IPAC Extragalactic Database
(NED) which is operated by the Jet Propulsion Laboratory, California
Institute of Technology, under contract with the National Aeronautics
and Space Administration. 
This research has
also made use of NASA's Astrophysics Data System Bibliographic
Service and {\sc asurv} Rev 1.2 \citep{lif92a}, which implements the 
methods presented in \citet{ifn86}.

%\bibliographystyle{apj}
%\bibliography{aa,ref}

\end{document}